\documentclass[12pt]{article}
\usepackage[dvips]{graphicx} % for figures
\usepackage{amsfonts}
\usepackage{amscd}
\usepackage{amsmath}    % need for subequations

\title{Practical Decoy State for Quantum Key Distribution}
\author{Xiongfeng Ma, Bing Qi, Yi Zhao, and Hoi-Kwong Lo\\
\\
\normalsize{Center for Quantum Information and Quantum Control} \\
\normalsize{Department of Physics and Department of Electrical \& Computer Engineering} \\
\normalsize{University of Toronto, Toronto,  Ontario, Canada} \\
}

\begin{document}

\maketitle

%%%%%%%%%%%%%%%%%%%%%%%%%%%%%%%%%%%%%%%%%%%%%%%%%%%%%%%%%%%%%%%%%%%%%%%
% Abstract
%%%%%%%%%%%%%%%%%%%%%%%%%%%%%%%%%%%%%%%%%%%%%%%%%%%%%%%%%%%%%%%%%%%%%%%
\begin{abstract}

Decoy states have recently been proposed as a useful method for
substantially improving the performance of quantum
key distribution. Here, we present a general theory of the
decoy state protocol based on only two decoy states and
one signal state. We perform optimization on the choice of
intensities of the two decoy states and the signal
state. Our result shows that a decoy state
protocol with only two types of decoy states----the
vacuum and a weak decoy state---asymptotically approaches the
theoretical limit of the most general type of decoy state
protocols (with an infinite
number of decoy states). We also
present a one-decoy-state protocol.
Moreover, we provide estimations on the effects
of statistical fluctuations and suggest that, even for
long distance (larger than 100km) QKD, our two-decoy-state
protocol can be implemented with only a few hours of experimental data.
In conclusion, decoy state quantum key distribution is highly practical.

\end{abstract}

%%%%%%%%%%%%%%%%%%%%%%%%%%%%%%%%%%%%%%%%%%%%%%%%%%%%%%%%%%%%%%%%%%%%%%%
% Abstract
%%%%%%%%%%%%%%%%%%%%%%%%%%%%%%%%%%%%%%%%%%%%%%%%%%%%%%%%%%%%%%%%%%%%%%%
\section{Introduction} \label{Intro}

The goal of quantum key distribution (QKD) \cite{BB84} is to allow
two distant parties, Alice and Bob, to share a common string of
secret (known as the key), in the presence of an eavesdropper, Eve.
Unlike conventional cryptography, QKD promises perfect security
based on the fundamental laws of physics. Proving the unconditional
security of QKD is a hard problem. Fortunately, this problem has
recently been solved \cite{Securityproofs,ShorPreskill}. See also
\cite{security}. Experimental QKD has been successfully demonstrated
over 100km of commercial Telecom fibers \cite{GYS,NEC} and
commercial QKD systems are already on the market. %\cite{MagiQidQuantique}.
The most important question of QKD is its
security. Real-life QKD systems are often based on attenuated laser
pulses (i.e., weak coherent states), which occasionally give out
more than one photon. This opens up the possibility of sophisticated
eavesdropping attacks such as a photon number splitting attack,
where Eve stops all single-photon signals and splits multi-photon
signals, keeping one copy herself and re-sending the rest to Bob.
The security of practical QKD systems has previously been discussed
in \cite{GLLP}.

Hwang \cite{HwangDecoy} proposed the decoy state method as an
important weapon to combat those sophisticated attack: by preparing
and testing the transmission properties of some decoy states, Alice
and Bob are in a much better position to catch an eavesdropper.
Hwang specifically proposed to use a decoy state with an average
number of photon of order $1$. Hwang's idea was highly innovative.
However, his security analysis was heuristic.

In \cite{Decoy}, we presented a rigorous security analysis of the
decoy state idea. More specifically, we combined the idea of the
entanglement
distillation approach in
GLLP\cite{GLLP} with the decoy method and achieved a formula for key
generation rate
\begin{equation} \label{Intro:KeyRate}
R \geq q \{-Q_{\mu}f(E_{\mu})H_2(E_{\mu})+Q_1[1-H_2(e_1)]\},
\end{equation}
where $q$ depends on the implementation (1/2 for the BB84 protocol
due to the fact that half of the time Alice and Bob disagree with
the bases, and if one uses the efficient BB84 protocol
\cite{EfficientBB84}, $q\approx1$), the subscript $\mu$ denotes
the intensity of signal states,
$Q_{\mu}$ is the gain \cite{gain} of signal
states, $E_{\mu}$ is the overall quantum bit error rate (QBER),
$Q_1$ is the gain of single photon states, $e_1$ is the error rate
of single photon states, $f(x)$ is the bi-direction error correction
efficiency (See, for example,
\cite{BS}.) as a function of error rate, normally $f(x)\ge1$ with
Shannon limit $f(x)=1$, and $H_2(x)$ is binary Shannon information
function, given by,
$$
H_2(x)=-x\log_2(x)-(1-x)\log_2(1-x).
$$

Four key variables are needed in Eq.~(\ref{Intro:KeyRate}).
$Q_{\mu}$ and $E_{\mu}$ can be measured directly from the
experiment. Therefore, in the paper \cite{Decoy}, we showed
rigorously how one can, using the decoy state idea to estimate $Q_1$
and $e_1$, thus achieving the unconditional security of QKD with the
key generation rate given by Eq.~(\ref{Intro:KeyRate}). Moreover,
using the experimental parameters from a particular QKD experiment
(GYS) \cite{GYS}, we showed that decoy state QKD can be secure over
140km of Telecom fibers. In summary, we showed clearly that decoy
state can indeed substantially increase both the distance and the
key generation rate of QKD.

For practical implementations, we \cite{Decoy} also emphasized that only a few
decoy states will be sufficient. This is so because contributions from
states with large photon numbers are negligible in comparison
with those from small photon numbers.
In particular, we proposed
a Vacuum+Weak decoy state protocol. That is to say, there are
two decoy states---a vacuum and a weak decoy state.
Moreover, the signal state is chosen to be of order $1$ photon on average.
The vacuum state is particularly useful for estimating
the background detection rate. Intuitively,
a weak decoy state allows us
to lower bound $Q_1$ and upper bound $e_1$.

Subsequently, the security of our
Vacuum+ Weak decoy state protocol has
been analyzed by Wang \cite{WangDecoy}.
Let us denote the intensities of the signal state and
the non-trivial decoy state by $\mu$ and $\mu'$ respectively.
Wang derived a useful upper bound for $\Delta$:
\begin{equation}\label{Intro:WangDelta}
\begin{aligned}
\Delta \le \frac{\mu}{\mu'-\mu}(\frac{\mu e^{-\mu}Q_{\mu'}}{\mu'
e^{-\mu'}Q_\mu}-1)+\frac{\mu e^{-\mu}Y_0}{\mu'Q_\mu},
\end{aligned}
\end{equation}
where $\Delta$ is the proportion of ``tagged'' states in the sifted
key as defined in GLLP \cite{GLLP}. Whereas we \cite{Decoy}
considered a strong version of GLLP result noted in
Eq.~\eqref{Intro:KeyRate}, Wang proposed to use a weak version of
GLLP result:
\begin{equation} \label{Intro:KeyRateWang}
R \geq q Q_{\mu}
\{-H_2(E_{\mu})+(1-\Delta)[1-H_2(\frac{E_\mu}{1-\Delta})]\}.
\end{equation}
Such a weak version of GLLP result does not require an estimation of
$e_1$. So, it has the advantage that the estimation process is
simple. However, it leads to lower values of the key generation
rates and distances. The issue of statistical fluctuations in decoy
state QKD was also mentioned in \cite{WangDecoy}.

Our observation \cite{Decoy} that only a few decoy
states are sufficient for practical implementations has been
studied further and confirmed in a recent paper
\cite{Harrington}, which is roughly concurrent to the present work.

The main goal of this paper is to analyze the security of a rather
general class of
two-decoy-state protocols with two weak decoy states and one signal state.
Our main contributions are as follows. First, we derive a general
theory for a decoy state protocol with two weak decoy states.
Whereas Wang \cite{WangDecoy}
considered only our Vacuum+Weak decoy state protocol\cite{Decoy}
(i.e., a protocol with two decoy states---the vacuum and a weak coherent
state),
our analysis here is more general. Our decoy method
applies even when both decoy states are non-vacuum. Note that,
in practice, it may
be difficult to prepare a vacuum decoy state. For instance,
standard VOAs (variable optical attenuators) cannot block
optical signals completely.
For the special case of the Vacuum+Weak decoy state protocol, our
result generalizes the work of Wang \cite{WangDecoy}.

Second, we perform an optimization of the key generation rate in
Eq.~\eqref{Intro:KeyRate} as a function of the intensities of the
two decoy states and the signal state. Up till now, such an
optimization problem has been a key unresolved problem in the
subject. We solve this problem analytically by showing that the key
generation rate given by Eq.~\eqref{Intro:KeyRate} is optimized when
both decoy states are weak. In fact, in the limit that both decoy
states are infinitesimally weak, we match the best lower bound on
$Y_1$ and upper bound of $e_1$ in the most general decoy state
theory where an infinite number of decoy states are used. Therefore,
asymptotically, there is no obvious advantage in using more than two
decoy states.

Third, for practical applications, we study the correction terms to
the key generation rate when the intensities of the two decoy states
are non-zero. We see that the correction terms (to the
asymptotically zero intensity case) are reasonably small.
For the case where one of the two decoy states is a vacuum (i.e.,
$v_2 =0$), the correction term remains modest even when
the intensity of the second decoy state, $\nu_1$ is as
high as $25\%$ of that of the
signal state.

Fourth, following \cite{WangDecoy}, we discuss the issue of statistical
fluctuations due to a
finite data size in real-life experiments. We provide a rough
estimation on the effects of statistical fluctuations in practical
implementations. Using a recent experiment \cite{GYS} as an example,
we estimate that, our weak decoy state proposal with two decoy
states (a vacuum and a weak decoy state of strength $\nu$) can
achieve secure QKD over more than 100km with only a few hours of
experiments. A caveat of our investigation is that we have not
considered the fluctuations in the intensities of Alice's laser
pulses (i.e., the values of $\mu, \nu_1$ and $ \nu_2$). This is
mainly because of a lack of reliable experimental data.
In summary, our result demonstrates that our two-decoy-state proposal is highly
practical.

Fifth, we also present a one-decoy-state protocol.
Such a protocol has an advantage of being simple to implement,
but gives a lower key generation rate.
Indeed, we have recently demonstrated experimentally
our one-decoy-state protocol over 15km \cite{experimentaldecoy}.
This demonstrates that one-decoy-state is, in fact, sufficient for
many practical applications.
In summary, decoy state QKD is simple and cheap to implement
and it is, therefore, ready for immediate commercialization.

We remark on passing that a different approach (based on strong
reference pulse) to making another protocol (B92 protocol) unconditionally
secure over a long distance has recently been proposed
in a theoretical paper by Koashi \cite{koashi}.

The organization of this paper is as follows. In section
\ref{Model}, we model an optical fiber based QKD set-up. In section
\ref{Decoy}, we first give a general
theory for $m$ decoy states.
We then propose our practical decoy method with two decoy
states. Besides, we optimize our choice of the average photon
numbers $\mu$ of the signal state and, $\nu_1$ and $\nu_2$
of the decoy states by
maximizing the key generation rate
with the experimental parameters in a specific QKD experiment
(GYS) \cite{GYS}.
Furthermore, we also present a simple one-decoy-state protocol.
In section \ref{Stab}, we discuss the effects of statistical
fluctuations in the two-decoy-state method for a finite data size
(i.e., the number of pulses transmitted by Alice).
Finally, in section \ref{s:Conclusion}, we present some concluding remarks.

%%%%%%%%%%%%%%%%%%%%%%%%%%%%%%%%%%%%%%%%%%%%%%%%%%%%%%%%%%%%%%%%%%%%%%%
% Model
%%%%%%%%%%%%%%%%%%%%%%%%%%%%%%%%%%%%%%%%%%%%%%%%%%%%%%%%%%%%%%%%%%%%%%%
\section{Model}\label{Model}
In order to describe a real-world QKD system, we need to model the
source, channel and detector. Here we consider a widely used fiber
based set-up model \cite{Lutkenhaus}. %p\&p

\textbf{Source:} The laser source can be modeled as a weak coherent
state. Assuming that the phase of each pulse is totally randomized,
the photon number of each pulse follows a Poisson distribution with
a parameter $\mu$ as its expected photon number set by Alice. Thus,
the density matrix of the state emitted by Alice is given by
\begin{equation}\label{Model:AliceState}
\rho_A=\sum^{\infty}_{i=0}\frac{\mu^i}{i!}\,e^{-\mu}\,
|i\rangle\langle i|,
\end{equation}
where $|0\rangle\langle 0|$ is vacuum state and $|i\rangle\langle
i|$ is the density matrix of i-photon state for $i=1,2\cdots$.

\textbf{Channel:} For optical fiber based QKD system, the losses in
the quantum channel can be derived from the loss coefficient
$\alpha$ measured in dB/km and the length of the fiber $l$ in km.
The channel transmittance $t_{AB}$ can be expressed as
$$
t_{AB}=10^{-\frac{\alpha l}{10}}.
$$

\textbf{Detector:} Let $\eta_{Bob}$ denote for the transmittance in
Bob's side, including the internal transmittance of optical
components $t_{Bob}$ and detector efficiency $\eta_D$,
$$
\eta_{Bob}=t_{Bob}\eta_D.
$$
Then the overall transmission and detection efficiency between Alice
and Bob $\eta$ is given by
\begin{equation}\label{Model:Eta}
\eta=t_{AB}\eta_{Bob}.
\end{equation}

It is common to consider a {\it threshold} detector in
Bob's side. That is to say, we assume that Bob's
detector can tell a vacuum from a non-vacuum state. However,
it cannot tell the actual photon number in the received signal,
if it contains at least one photon.

It is reasonable to assume the independence between the behaviors of
the $i$ photons in i-photon states. Therefore the transmittance of
\emph{i-photon} state $\eta_i$ with respect to a threshold detector
is given by
\begin{equation}\label{Model:etai}
\eta_i=1-(1-\eta)^i
\end{equation}
for $i=0,1,2,\cdots$.

\textbf{Yield:} define $Y_i$ to be the yield of an i-photon state,
i.e., the conditional probability of a detection event at Bob's side
given that Alice sends out an i-photon state. Note that $Y_0$ is the
background rate which includes the detector dark count and other
background contributions such as the stray light from timing pulses.

The yield of i-photon states $Y_i$ mainly come from two parts,
background and true signal. Assuming that the background counts are
independent of the signal photon detection, then $Y_i$ is given by
\begin{equation}\label{Model:Yi}
\begin{aligned}
Y_i &= Y_0 + \eta_i - Y_0\eta_i \\
    &\cong Y_0 + \eta_i.
\end{aligned}
\end{equation}
Here we assume $Y_0$ (typically $10^{-5}$) and $\eta$ (typically
$10^{-3}$) are small.

The {\it gain} of i-photon states $Q_i$ is given by
\begin{equation}\label{Model:Qi}
\begin{aligned}
Q_i &= Y_i\frac{\mu^i}{i!}e^{-\mu}.
\end{aligned}
\end{equation}
The gain $Q_i$ is the product of the probability of Alice sending
out an $i$-photon state (follows Poisson distribution) and the
conditional probability of Alice's $i$-photon state (and background)
will lead to a detection event in Bob.

\textbf{Quantum Bit Error Rate:}
The error rate of i-photon states $e_i$ is given by
\begin{equation}\label{Model:ei}
e_i = \frac{e_0 Y_0 + e_{detector}\eta_i}{Y_i}
\end{equation}
where $e_{detector}$ is the probability that a photon hit the
erroneous detector. $e_{detector}$ characterizes the alignment and
stability of the optical system. Experimentally, even
at distances as long as 122km,
$e_{detector}$ is more or less independent of the distance.
In what follows, we will assume that $e_{detector}$ is a constant.
We will assume that the background
is random. Thus the error rate of the background is $e_0=\frac12$.
Note that Eqs.~\eqref{Model:etai}, \eqref{Model:Yi},
\eqref{Model:Qi} and \eqref{Model:ei} are satisfied for all
$i=0,1,2,\cdots$.

The overall gain is given by
\begin{equation}\label{Model:Gain}
\begin{aligned}
Q_{\mu} &= \sum_{i=0}^{\infty} Y_i\frac{\mu^i}{i!}e^{-\mu} \\
           &= Y_0 + 1-e^{-\eta\mu}.
\end{aligned}
\end{equation}

The overall QBER is given by
\begin{equation}\label{Model:QBER}
\begin{aligned}
E_{\mu}Q_{\mu} &= \sum_{i=0}^{\infty} e_iY_i\frac{\mu^i}{i!}e^{-\mu} \\
                     &= e_0 Y_0 + e_{detector}(1-e^{-\eta\mu}).
\end{aligned}
\end{equation}

%%%%%%%%%%%%%%%%%%%%%%%%%%%%%%%%%%%%%%%%%%%%%%%%%%%%%%%%%%%%%%%%%%%%%%%
% Decoy
%%%%%%%%%%%%%%%%%%%%%%%%%%%%%%%%%%%%%%%%%%%%%%%%%%%%%%%%%%%%%%%%%%%%%%%
\section{Practical decoy method} \label{Decoy}
In this section, we will first discuss the choice of $\mu$ for the
signal state to maximize the key generation rate as given by
Eq.~\eqref{Intro:KeyRate}. Then, we will consider a specific
protocol of two weak decoy states and show how they can be used to
estimate $Y_1$ and $e_1$ rather accurately. After that, we will show
how to choose two decoy states to optimize the key generation rate
in Eq.~\eqref{Intro:KeyRate}. As a whole, we have a practical decoy
state protocol with two weak decoy states.

%%%%%%%%%%%%%%%%%%%%%%%%%%%%%%%%%%%%%%%%%%%%%%%%%%%%%%%%%%%%%%%%%%%%%%%
% OptMu
%%%%%%%%%%%%%%%%%%%%%%%%%%%%%%%%%%%%%%%%%%%%%%%%%%%%%%%%%%%%%%%%%%%%%%%
\subsection{Choose optimal $\mu$} \label{OptMu}
Here we will discuss how to choose the expected photon number of
signal states $\mu$ to maximize the key generation rate in
Eq.~\eqref{Intro:KeyRate}.

Let us begin with a general discussion. On one hand, we need to
maximize the gain of single photon state $Q_1$, which is the only
source for the final secure key. To achieve this, heuristically, we
should maximize the probability of Alice sending out single photon
signals. With a Poisson distribution of the photon number, the
single photon fraction in the signal source reaches its maximum when
$\mu=1$. On the other hand, we have to control the gain of multi
photon state to ensure the security of the system. Thus, we should
keep the fraction ${Q_1}/{Q_\mu}$ high,
which requires $\mu$ not to be too large. %Based on both points, the case
%of $\mu>1$ is always worse than the case of $\mu=1$.
Therefore,
intuitively we have
$$
\mu \in (0,1].
$$

As will be noted in the next Subsection, Alice and Bob can estimate
$e_1$ and $Y_1$ rather accurately in a simple decoy state protocol
(e.g., one involving only two decoy states). Therefore, for ease of
discussion, we will discuss the case where Alice and Bob can
estimate $e_1$ and $Y_1$ perfectly. Minor errors in Alice and Bob's
estimation of $e_1$ and $Y_1$ will generally lead to rather modest
change to the final key generation rate $R$. According to
Eqs.~\eqref{Model:Qi} and \eqref{Model:ei}, $Q_1$ will be maximized
when $\mu=1$ and $e_1$ is independent of $\mu$, so we can expect
that the optimal expected photon number of signal state is
$\mu=O(1)$.
%
%The key generation rate will be maximized when
%\begin{equation} \label{OptMu:General}
%\frac{\partial R}{\partial\mu}=0.
%\end{equation}

We consider the case where the background rate is low ($Y_0\ll\eta$)
and the transmittance is small $\eta\ll1$ (typical values:
$Y_0=10^{-5}$ and $\eta=10^{-3}$). By substituting
Eqs.~\eqref{Model:Qi}, \eqref{Model:ei}, \eqref{Model:Gain} and
\eqref{Model:QBER} into Eq.~\eqref{Intro:KeyRate}, the key
generation rate is given by,
$$
R \approx -\eta\mu f(e_{detector})H_2(e_{detector})+\eta\mu
e^{-\mu}[1-H_2(e_{detector})]
$$
This rate is optimized if we choose $\mu=\mu_{optimal}$ which
fulfills,
\begin{equation}\label{OptMu:Final}
\begin{aligned}
(1-\mu)\exp(-\mu)=\frac{f(e_{detector})H_2(e_{detector})}{1-H_2(e_{detector})},
\end{aligned}
\end{equation}
where $e_{detector}$ is the probability that a photon hits the
erroneous detector. Then, using the data shown in Table
\ref{OptMu:Tab:Data} extracted from a recent experiment \cite{GYS},
we can solve this equation and obtain that, $\mu_{optimal}^{GYS}
\approx 0.54$ for $f(e)=1$ and $\mu_{optimal}^{GYS} \approx 0.48$
for $f(e)=1.22$. As noted in \cite{Decoy}, the key generation rate
and distance are pretty stable against even a $20\%$ change of
$\mu$.

%After comparing with numerical analysis of Eq.
%\eqref{OptMu:General}, we can see that Eq.~\eqref{OptMu:Final} is a
%good approximation.
\begin{table}[htb]\center
%\newcounter{Table}
\begin{tabular}{|c|c|c|c|c|c|c|}
\hline
Experiment & $\lambda$ [nm] & $\alpha$ [dB/km] & $e_{detector}$ [\%] & $Y_0$ & $\eta_{Bob}$ & $f$ \\
\hline
GYS~\cite{GYS} & 1550 & 0.21 & 3.3 & $1.7\times10^{-6}$ & 0.045 & $2MHz$ \\
\hline
KTH~\cite{KTH} & 1550 & 0.2 & 1 & $4\times10^{-4}$ & 0.143 & $0.1MHz$ \\
\hline
\end{tabular}
%\caption{\normalfont{data from sth to say, sth to hehe}}
\caption{Key parameters for QKD experiments.} \label{OptMu:Tab:Data}
\end{table}

\subsection{General decoy method} \label{General}
Here we will give out the most general decoy state method with $m$
decoy states. This extends our earlier work in \cite{Decoy}.

Suppose Alice and Bob choose the signal and decoy states with
expected photon number $\mu,\nu_1,\nu_2,\cdots,\nu_m$, they will get
the gains and QBER's for signal state and decoy states,
\begin{equation}\label{Decoy:SigDecoym}
\begin{aligned}
Q_{\mu}e^{\mu} &= \sum_{i=0}^{\infty} Y_i\frac{\mu^i}{i!} \\
E_{\mu}Q_{\mu}e^{\mu} &= \sum_{i=0}^{\infty} e_iY_i\frac{\mu^i}{i!} \\
Q_{\nu_1}e^{\nu_1} &= \sum_{i=0}^{\infty}Y_i\frac{\nu_1^i}{i!}  \\
E_{\nu_1} Q_{\nu_1} e^{\nu_1} &= \sum_{i=0}^{\infty}e_iY_i\frac{\nu_1^i}{i!} \\
Q_{\nu_2}e^{\nu_2} &= \sum_{i=0}^{\infty}Y_i\frac{\nu_2^i}{i!}  \\
E_{\nu_2} Q_{\nu_2} e^{\nu_2} &= \sum_{i=0}^{\infty}e_iY_i\frac{\nu_2^i}{i!} \\
\cdots \\
Q_{\nu_m}e^{\nu_m} &= \sum_{i=0}^{\infty}Y_i\frac{\nu_m^i}{i!}  \\
E_{\nu_m} Q_{\nu_m} e^{\nu_m} &= \sum_{i=0}^{\infty}e_iY_i\frac{\nu_m^i}{i!} \\
\end{aligned}
\end{equation}
\textbf{Question:} given Eqs.~\eqref{Decoy:SigDecoym}, how can one
find a tight lower bound of $R$, which is given by Eq.~\eqref{Intro:KeyRate}?
This is a main optimization problem for the design of decoy state protocols.
%Here in Eq.~\eqref{Intro:KeyRate}, all
%$\{Y_i\}$ and $\{e_i\}$ are unknown for $i=0,1,2,\cdots$. Here the
%restrictions Eqs.~\eqref{Decoy:ConditionWeak} are not necessary.

Note that in Eq.~\eqref{Intro:KeyRate}, the first term and $q$ are
independent of $\{Y_i\}$ and $\{e_i\}$.% (here we haven't take the
%statistical fluctuation into account).
Combining with Eq.~\eqref{Model:Qi}, we can simplify the problem to:

\textit{How to lower bound}
\begin{equation}\label{Decoy:Priv}
\begin{aligned}
P = Y_1[1-H_2(e_1)] \\
\end{aligned}
\end{equation}
\textit{with the constraints given by Eqs.~\eqref{Decoy:SigDecoym}?}

When $m\rightarrow\infty$, Alice and Bob can solve all $\{Y_i\}$ and
$\{e_i\}$ accurately in principle. This is the asymptotic case given
in \cite{Decoy}.

\subsection{Two decoy states} \label{TwoDecoy}
As emphasized in \cite{Decoy}, only a few decoy states are needed
for practical implementations. A simple way to lower bound
Eq.~\eqref{Decoy:Priv} is to lower bound $Y_1$ and upper bound
$e_1$. Intuitively, only two decoy states are needed for the
estimation of $Y_1$ and $e_1$ and, therefore, for
practical decoy state implementation. %A vacuum decoy state will
%allow us to estimate the background rate whereas a weak decoy state
%will allow us to lower bound $Y_1$ and $e_1$ for
%Eq.~\eqref{Intro:KeyRate}.
Here, we present a rigorous analysis to show more precisely how to
use two weak decoy states to estimate the lower bound $Y_1$ and
upper bound $e_1$.

Suppose Alice and Bob choose two decoy states with expected photon
numbers $\nu_1$ and $\nu_2$ which satisfy
\begin{equation}\label{Decoy:ConditionWeak}
\begin{aligned}
%\nu_1>\nu_2\ge0\\
0\le\nu_2<\nu_1\\
\nu_1+\nu_2<\mu,
\end{aligned}
\end{equation}
where $\mu$ is the expected photon number of the signal state.

\textbf{Lower bound of $Y_1$:} Similar to Eq.~\eqref{Model:Gain},
the gains of these two decoy states are given by
\begin{eqnarray}
%\begin{equation}\label{Decoy:WeGain}
%\begin{aligned}
Q_{\nu_1} &=& \sum_{i=0}^{\infty}Y_i\frac{\nu_1^i}{i!} e^{-\nu_1}
\label{Decoy:WeGain1}\\
Q_{\nu_2} &=& \sum_{i=0}^{\infty}Y_i\frac{\nu_2^i}{i!} e^{-\nu_2}
\label{Decoy:WeGain2}
%\end{aligned}
%\end{equation}
\end{eqnarray}

First Alice and Bob can estimate the lower bound of background rate
$Y_0$ by
$\nu_1\times\eqref{Decoy:WeGain1}-\nu_2\times\eqref{Decoy:WeGain2}$,
%$\nu_1Q_{\nu_2}e^{\nu_2}-\nu_2Q_{\nu_1}e^{\nu_1}$ from
%Eqs.~\eqref{Decoy:WeGain1} and \eqref{Decoy:WeGain2},
$$
\begin{aligned}
\nu_1Q_{\nu_2}e^{\nu_2}-\nu_2Q_{\nu_1}e^{\nu_1} &=
(\nu_1-\nu_2)Y_0-\nu_1\nu_2(Y_2\frac{\nu_1-\nu_2}{2!}
+Y_3\frac{\nu_1^2-\nu_2^2}{3!}+\cdots) \\
&\le (\nu_1-\nu_2)Y_0.
\end{aligned}
$$
Thus, a crude lower bound of $Y_0$ is given by
\begin{equation}\label{Decoy:Y0Low}
\begin{aligned}
Y_0 \ge Y_0^L = \max\{
\frac{\nu_1Q_{\nu_2}e^{\nu_2}-\nu_2Q_{\nu_1}e^{\nu_1}}{\nu_1-\nu_2},0\},
\end{aligned}
\end{equation}
where the equality sign will hold when $\nu_2=0$, that is to say,
when a vacuum decoy ($\nu_2=0$) is performed,
Eq.~\eqref{Decoy:Y0Low} is tight.

Now, from Eq.~\eqref{Model:Gain}, the contribution from multi photon
states (with photon number $\ge2$) in signal state can be expressed
by,
\begin{equation}\label{Decoy:MulGain}
\begin{aligned}
\sum_{i=2}^{\infty}Y_i\frac{\mu^i}{i!} &= Q_\mu e^\mu-Y_0-Y_1\mu\\
\end{aligned}
\end{equation}

Combining Eqs.~\eqref{Decoy:WeGain1} and \eqref{Decoy:WeGain2},
under condition Eq.~\eqref{Decoy:ConditionWeak}, we have
\begin{equation}\label{Decoy:WeGainBound}
\begin{aligned}
Q_{\nu_1}e^{\nu_1}-Q_{\nu_2}e^{\nu_2} &= Y_1(\nu_1-\nu_2) + \sum_{i=2}^{\infty}\frac{Y_i}{i!}(\nu_1^i-\nu_2^i)\\
%&= Y_0 + Y_1\nu + \sum_{i=2}^{\infty}Y_i\frac{\nu^{i}}{i!}\frac{\mu^{i}}{\mu^{i}}\\
&\le Y_1(\nu_1-\nu_2) + \frac{\nu_1^{2}-\nu_2^2}{\mu^{2}}\sum_{i=2}^{\infty}Y_i\frac{\mu^{i}}{i!}\\
&= Y_1(\nu_1-\nu_2) + \frac{\nu_1^{2}-\nu_2^2}{\mu^{2}}(Q_\mu e^\mu-Y_0-Y_1\mu)\\
&\le Y_1(\nu_1-\nu_2) + \frac{\nu_1^{2}-\nu_2^2}{\mu^{2}}(Q_\mu e^\mu-Y_0^L-Y_1\mu),\\
\end{aligned}
\end{equation}
where $Y_0^L$ was defined in Eq.~\ref{Decoy:Y0Low}.
Here, to prove the first inequality in Eq~\eqref{Decoy:WeGainBound},
we have made use of the inequality that $a^i - b^i \le a^2 - b^2$
whenever $0<a + b <1 $ and $i \ge 2$. The equality sign holds for
the first inequality in Eq~\eqref{Decoy:WeGainBound} if and only if
Eve raises the yield of 2-photon states and blocks all the states
with photon number greater than $2$ (This was also mentioned in
\cite{HwangDecoy}). The second equality in
Eq~\eqref{Decoy:WeGainBound} is due to Eq.~\eqref{Decoy:Y0Low}. %the fact that Alice and Bob
%do not know the background rate, $Y_0$, in general. Actually Alice
%and Bob can use two weak decoy states to estimate the lower bound of
%$Y_0$, which will give tighter bound of $Y_1$, discussed in
%Appendix~\ref{Better2decoy}. If a vacuum decoy state is used, then
%the second inequality will be not necessary.

By solving inequality \eqref{Decoy:WeGainBound}, the lower bound of
$Y_1$ is given by
\begin{equation}\label{Decoy:Y1Bound}
\begin{aligned}
Y_1 \ge Y_1^{L,\nu_1,\nu_2} =
\frac{\mu}{\mu\nu_1-\mu\nu_2-\nu_1^2+\nu_2^2}
[Q_{\nu_1}e^{\nu_1}-Q_{\nu_2}e^{\nu_2}-\frac{\nu_1^2-\nu_2^2}{\mu^2}(Q_\mu
e^\mu-Y_0^L)].
\end{aligned}
\end{equation}
Then the gain of single photon state is given by, according to
Eq.~\eqref{Model:Qi},
\begin{equation}\label{Decoy:Q1Bound}
\begin{aligned}
Q_1 \ge Q_1^{L,\nu_1,\nu_2} =
\frac{\mu^2e^{-\mu}}{\mu\nu_1-\mu\nu_2-\nu_1^2+\nu_2^2}
[Q_{\nu_1}e^{\nu_1}-Q_{\nu_2}e^{\nu_2}-\frac{\nu_1^2-\nu_2^2}{\mu^2}(Q_\mu
e^\mu-Y_0^L)],
\end{aligned}
\end{equation}
where $Y_0^L$ is given by Eq.~\eqref{Decoy:Y0Low}.

\textbf{Upper bound of $e_1$:} According to Eq.~\eqref{Model:QBER},
the QBER of the weak decoy state is given by
\begin{eqnarray}
%\begin{equation}\label{Decoy:nuQBER}
%\begin{aligned}
E_{\nu_1} Q_{\nu_1} e^{\nu_1} &=& e_0Y_0+e_1\nu_1Y_1+\sum_{i=2}^{\infty}e_iY_i\frac{\nu_1^i}{i!}\label{Decoy:WeQBER1}\\
E_{\nu_2} Q_{\nu_2} e^{\nu_2} &=&
e_0Y_0+e_1\nu_2Y_1+\sum_{i=2}^{\infty}e_iY_i\frac{\nu_2^i}{i!}\label{Decoy:WeQBER2}
%\end{aligned}
%\end{equation}
\end{eqnarray}
An upper bound of $e_1$ can be obtained directly from
Eqs.~\eqref{Decoy:WeQBER1}-\eqref{Decoy:WeQBER2},
\begin{equation}\label{Decoy:e1Bound}
\begin{aligned}
e_1 \le e_1^{U,\nu_1,\nu_2} = \frac{E_{\nu_1}
Q_{\nu_1}e^{\nu_1}-E_{\nu_2} Q_{\nu_2}
e^{\nu_2}}{(\nu_1-\nu_2)Y_1^{L,\nu_1,\nu_2}}.
\end{aligned}
\end{equation}
Note that Alice and Bob should substitute the lower bound of
$Y_1$, Eq.~\eqref{Decoy:Y1Bound} into Eq.~\eqref{Decoy:e1Bound} to get
an upper bound of $e_1$.

In summary, by using two weak decoy states that satisfy
Eq.~\eqref{Decoy:ConditionWeak}, Alice and Bob can obtain a lower
bound for the yield $Y_1$ with Eq.~\eqref{Decoy:Y1Bound} (and then
the gain $Q_1$ with Eq.~\eqref{Decoy:Q1Bound}) and an upper bound
for the QBER $e_1$ with Eq.~\eqref{Decoy:e1Bound} for the single
photon signals. Subsequently, they can use Eq.~\eqref{Intro:KeyRate}
to work out the key generation rate as
\begin{equation}\label{TwodecoyKeyRate}
R \geq q \{-Q_{\mu}f(E_{\mu})H_2(E_{\mu})+
Q_1^{L,\nu_1,\nu_2}[1-H_2( e_1^{U,\nu_1,\nu_2}  )]\}.
\end{equation}

This is the main procedure of
our two-decoy-state protocol.

Now, the next question is: How good are our bounds for $Y_1$ and $e_1$
for our two-decoy-state protocol?
In what follows, we will examine the performance of our two weak
decoy state protocol by considering first the asymptotic case
where both $\nu_1$ and $\nu_2$ tend to $0$.
We will show that our bounds for $Y_1$ and $e_1$ are tight
in this asymptotic limit.

\textbf{Asymptotic case:} We will now take the limit $\nu_1 \to 0$
and $\nu_2 \to 0$. When $\nu_2<\nu_1\ll\mu=O(1)$, substituting
Eqs.~\eqref{Model:Gain}, \eqref{Decoy:WeGain1} and
\eqref{Decoy:WeGain2} into Eq. \eqref{Decoy:Y1Bound}, the lower
bound of $Y_1$ becomes
\begin{equation}\label{Decoy:Y1BoundAppro}
\begin{aligned}
Y_1^{L,0} &= Y_1^{L,\nu_1,\nu_2}\mid_{\nu_1 \rightarrow 0,\nu_2\rightarrow0} \\
          &= \frac{\mu}{\mu\nu_1-\mu\nu_2-\nu_1^2+\nu_2^2}(Q_{\nu_1}e^{\nu_1}-Q_{\nu_2}e^{\nu_2})\mid_{\nu_1 \rightarrow 0,\nu_2\rightarrow0} \\
          &= \frac{\mu}{\mu-\nu_1-\nu_2}\cdot\frac{1}{\nu_1-\nu_2}[(Y_0+\eta\nu_1)e^{\nu_1}-(Y_0+\eta\nu_2)e^{\nu_2}]\mid_{\nu_1 \rightarrow 0,\nu_2\rightarrow0} \\
          &= Y_0+\eta,
\end{aligned}
\end{equation}
which matches the theoretical value $Y_1\cong Y_0+\eta$ from
Eq.~\eqref{Model:Yi}. Substituting Eqs.~\eqref{Model:QBER},
\eqref{Decoy:WeQBER1} and \eqref{Decoy:WeQBER2} into Eq.
\eqref{Decoy:e1Bound}, the upper bound of $e_1$ becomes
\begin{equation}\label{Decoy:e1BoundAppro}
\begin{aligned}
e_1^{U,0} &= e_1^{U,\nu_1,\nu_2}\mid_{\nu_1 \rightarrow0, \nu_2 \rightarrow0} \\
    &= \frac{e_0Y_0+e_{detector}\eta}{Y_1}, \\
%    &\cong \eta+Y_0,
\end{aligned}
\end{equation}
which matches the theoretical value from Eq.~\eqref{Model:ei}.

The above calculation seems to suggest that our two-decoy-state
protocol is as good as the most general protocol in the limit
$\nu_1, \nu_2 \to 0$. However, in real-life, at least one of the two
quantities $\nu_1$ and $\nu_2$ must take on a non-zero value.
Therefore, we need to study the effects of finite $\nu_1$ and
$\nu_2$. This will be our next subject.

\textbf{Deviation from theoretical values:} Here, we consider how
finite values of $\nu_1$ and perhaps $\nu_2$ will change our bounds for $Y_1$ and
$e_1$.

The relative deviation
of $Y_1$ is given by
\begin{equation}\label{Decoy:Y1Dev}
\begin{aligned}
\beta_{Y1} &= \frac{Y_1^{L,0}-Y_1^{L,\nu_1,\nu_2}}{Y_1^{L,0}}, \\
%&= [Y_0+\eta-\frac{\mu}{\mu\nu-\nu^2}(Q_\nu
%e^{\nu}-Q_\mu e^\mu\frac{\nu^2}{\mu^2}-\frac{\mu^2-\nu^2}{\mu^2}Y_0)]/Y_1 \\
%&\approx 1-\frac{\mu}{\mu\nu-\nu^2}(\nu e^{\nu}-\mu
%e^\mu\frac{\nu^2}{\mu^2})\\
%&\approx \frac\nu{\mu-\nu}(e^\mu-1-\mu)\cdot\eta +
%\frac\nu{\mu(\mu-\nu)}(e^\mu-1-\mu-\frac{\mu^2}2)\cdot Y_0. \\
%&\approx \frac\nu{\mu}(e^\mu-1-\mu)\cdot\eta +
%\frac\nu{\mu^2}(e^\mu-1-\mu-\frac{\mu^2}2)\cdot Y_0. \\
%&\approx (e^\mu-1-\mu)\frac\nu{\mu} - \frac{Y_0}{2Y_1}\nu. \\
\end{aligned}
\end{equation}
where $Y_1^{L,0}$ is the theoretical value of $Y_1$ given in
Eqs.~\eqref{Model:Yi} and \eqref{Decoy:Y1BoundAppro}, and
$Y_1^{L,\nu_1,\nu_2}$ is an estimation value of $Y_1$ by our
two--decoy-state method as given in Eq.~\eqref{Decoy:Y1Bound}.

The relative deviation of $e_1$ is given by
\begin{equation}\label{Decoy:e1Dev}
\begin{aligned}
\beta_{e1} &= \frac{e_1^{U,\nu_1,\nu_2}-e_1^{U,0}}{e_1^{U,0}}, \\
%&\le [\frac{E_\nu Q_\nu e^{\nu}-e_0Y_0}{(1-\beta_{Y1})Y_1\nu}-e_1]/e_1 \\
%&\approx \frac{e_1Y_1\nu+e_2Y_2\nu^2/2}{e_1Y_1\nu}+\beta_{Y1}-1 \\
%&\approx \frac{e_2Y_2\nu^2/2}{e_1Y_1\nu}+\beta_{Y1} \\
%&\approx (e^\mu-1)\frac\nu\mu-\frac{Y_0}{2Y_1}\nu. \\
%&\approx \frac\nu{\mu}(e^\mu-1-\mu)\cdot\eta +
%\frac\nu{\mu^2}(e^\mu-1-\mu-\frac{\mu^2}2)\cdot Y_0. \\
%&\approx \nu+\beta_{Y1} \\
\end{aligned}
\end{equation}
where $e_1^{L,0}$ is the theoretical value of $e_1$ given in
Eqs.~\eqref{Model:ei} and \eqref{Decoy:e1BoundAppro}, and
$e_1^{L,\nu_1,\nu_2}$ is the estimation value of $e_1$ by our
two-decoy-state method as given in Eq.~\eqref{Decoy:e1Bound}.

Under the approximation $\eta\ll1$ and taking the first order in
$\nu_1$ and $\nu_2$, and substituting Eqs.~\eqref{Model:Yi},
\eqref{Model:Gain}, \eqref{Decoy:WeGain1}, \eqref{Decoy:WeGain2},
\eqref{Decoy:Y0Low} and \eqref{Decoy:Y1Bound} into
Eq.~\eqref{Decoy:Y1Dev}, the deviation of the lower bound of $Y_1$
is given by
\begin{equation}\label{Decoy:Y1DevAppro}
\begin{aligned}
Y_1 \beta_{Y1}
&= Y_1^{L,0}-Y_1^{L,\nu_1,\nu_2} \\
&= Y_0+\eta-\frac{\mu}{\mu\nu_1-\mu\nu_2-\nu_1^2+\nu_2^2}
[Q_{\nu_1}e^{\nu_1}-Q_{\nu_2}e^{\nu_2}-\frac{\nu_1^2-\nu_2^2}{\mu^2}(Q_\mu
e^\mu-Y_0^L)] \\
%&\approx 1-\frac{\mu}{\mu\nu-\nu^2}(\nu e^{\nu}-\mu
%e^\mu\frac{\nu^2}{\mu^2})\\
%&\approx \frac\nu{\mu-\nu}(e^\mu-1-\mu)\cdot\eta +
%\frac\nu{\mu(\mu-\nu)}(e^\mu-1-\mu-\frac{\mu^2}2)\cdot Y_0. \\
%%%%%%%%%%%%%%%%%%
%&\approx Y_0+\eta-\frac{\mu}{\mu\nu_1-\mu\nu_2-\nu_1^2+\nu_2^2}
%[(Y_0+\eta\nu_1)e^{\nu_1}-(Y_0+\eta\nu_2)e^{\nu_2}-\frac{\nu_1^2-\nu_2^2}{\mu^2}(Y_0+\eta\mu)e^\mu] \\
%%%%%%%%%%%%%%%%%%
%&\approx (e^\mu-1-\mu)\frac\nu{\mu} - \frac{Y_0}{2Y_1}\nu. \\
&\approx
(e^\mu-1-\mu-\frac{\mu^2}{2})(\frac{1}{\mu-\nu_1-\nu_2}-\frac{1}{\mu})\cdot Y_0 + (e^\mu-1-\mu)\frac{\nu_1+\nu_2}{\mu-\nu_1-\nu_2}\cdot \eta.\\
\end{aligned}
\end{equation}
Substituting Eqs.~\eqref{Model:ei}, \eqref{Model:QBER},
\eqref{Decoy:WeQBER1}, \eqref{Decoy:WeQBER2}, \eqref{Decoy:e1Bound}
and \eqref{Decoy:Y1DevAppro} into Eq. \eqref{Decoy:e1Dev}, the
deviation of the upper bound of $e_1$ is given by
\begin{equation}\label{Decoy:e1DevAppro}
\begin{aligned}
e_1\beta_{e1} &= e_1^{U,\nu,0}-e_1^{U,0}\\
%%%%%%%%%%%%%%%%%%%%%%%%%%
%&= \frac{E_{\nu_1} Q_{\nu_1} e^{\nu_1}-E_{\nu_2} Q_{\nu_2} e^{\nu_2}}{(\nu_1-\nu_2)(1-\beta_{Y1})Y_1}-e_1 \\
%&= \frac{[e_0Y_0+e_{detector}(1-e^{-\eta\nu_1})] e^{\nu_1} - [e_0Y_0+e_{detector}(1-e^{-\eta\nu_2})] e^{\nu_2}}{(\nu_1-\nu_2)(1-\beta_{Y1})Y_1}-e_1 \\
%&\approx \frac{[1+\frac12(\nu_1+\nu_2)]e_0Y_0+(1+\nu_1+\nu_2)e_{detector}\eta}{(1-\beta_{Y1})Y_1}-e_1 \\
%&\approx e_1\beta_{Y1}+\frac{\frac12(\nu_1+\nu_2)e_0Y_0+(\nu_1+\nu_2)e_{detector}\eta}{Y_0+\eta} \\
%%%%%%%%%%%%%%%%%%%%%%%%%%
&= e_1\beta_{Y1}+(\nu_1+\nu_2)(e_1-\frac{e_0Y_0}{2Y_1}). \\
%&\approx \frac{e_1Y_1\nu+e_2Y_2\nu^2/2}{e_1Y_1\nu}+\beta_{Y1}-1 \\
%&\approx \frac{e_2Y_2\nu^2/2}{e_1Y_1\nu}+\beta_{Y1} \\
%&\approx (e^\mu-1)\frac\nu\mu-\frac{Y_0}{2Y_1}\nu. \\
%&\approx \frac\nu{\mu}(e^\mu-1-\mu)\cdot\eta +
%\frac\nu{\mu^2}(e^\mu-1-\mu-\frac{\mu^2}2)\cdot Y_0. \\
%&\approx e_1\beta_{Y1}+\nu, \\
\end{aligned}
\end{equation}

Now, from Eqs.~\eqref{Decoy:Y1DevAppro} and
\eqref{Decoy:e1DevAppro}, we can see that decreasing $\nu_1+\nu_2$
will improve the estimation of $Y_1$ and $e_1$. So, the smaller
$\nu_1+\nu_2$ is, the higher the key generation rate $R$ is. In
Appendix A, we will prove that decreasing $\nu_1+\nu_2$ will improve
the estimation of $Y_1$ and $e_1$ in general sense (i.e., without
the limit $\eta\ll1$ and taking the first order in $\nu_1$ and
$\nu_2$). Therefore, we have reached the following important
conclusion: for any fixed value of $\nu_1$, the choice $\nu_2=0$
will optimize the key generation rate. In this sense, the
Vacuum+Weak decoy state protocol, as first proposed in
an intuitive manner in \cite{Decoy},
is, in fact, optimal.

The above conclusion highlights the importance of the Vacuum+Weak
decoy state protocol. We will discuss them in following subsection.
Nonetheless, as remarked earlier, in practice, it might not be easy
to prepare a true vacuum state (with say VOAs). Therefore, our
general theory on non-zero decoy states, presented in this
subsection, is important.

\subsection{Vacuum+Weak decoy state} \label{V+W}
Here we will introduce a special case of Subsection~\ref{TwoDecoy}
with two decoy states: vacuum and weak decoy state. This special
case was first proposed in \cite{Decoy} and analyzed in
\cite{WangDecoy}. In the end of Subsection~\ref{TwoDecoy}, we have
pointed out that this case is optimal for two-decoy-state method.

\textbf{Vacuum decoy state:} Alice shuts off her photon source to
perform vacuum decoy state. Through this decoy state, Alice and Bob
can estimate the background rate,
%$$
\begin{equation}\label{Decoy:VaQE}
\begin{aligned}
Q_{vacuum} &= Y_0 \\
E_{vacuum} &= e_0 = \frac12.
\end{aligned}
\end{equation}
%$$
The dark counts occur randomly, thus the error rate of dark count is
$e_0=\frac12$.

\textbf{Weak decoy state:} Alice and Bob choose a relatively weak
decoy state with expected photon number $\nu<\mu$. %The gain and QBER
%will given by
%\begin{equation}\label{Decoy:WeQE}
%\begin{aligned}
%Q_\nu &= \sum_0^{\infty}{Y_i\frac{\nu^i}{i!}} \\
%E_\nu Q_\nu &= \sum_0^{\infty}{e_iY_i\frac{\nu^i}{i!}}. \\
%\end{aligned}
%\end{equation}

Here is the key difference between this special case and our general
case of two-decoy-state protocol. Now, from vacuum decoy state,
Eq.~\eqref{Decoy:VaQE}, Alice and Bob can estimate $Y_0$ accurately.
So, the second inequality of Eq.~\eqref{Decoy:WeGainBound} will be tight. %is {\it
%not} necessary.
%This is also discussed in Appendix~\ref{Better2decoy}.
Similar to Eq.~\eqref{Decoy:Y1Bound}, the lower bound of $Y_1$ is
given by
%\begin{eqnarray}\label{Decoy:WeGain}
\begin{equation}\label{Decoy:VWY1Bound}
\begin{aligned}
Y_1 \ge Y_1^{L,\nu,0} = Y_1^{L,\nu,\nu_2}\mid_{\nu_2\rightarrow0} =
\frac{\mu}{\mu\nu-\nu^2}(Q_\nu e^{\nu}-Q_\mu
e^\mu\frac{\nu^2}{\mu^2}-\frac{\mu^2-\nu^2}{\mu^2}Y_0).
\end{aligned}
\end{equation}
%\end{eqnarray}

So the gain of single photon state is given by,
Eq.~\eqref{Model:Qi},
\begin{equation}\label{Decoy:VWQ1Bound}
\begin{aligned}
Q_1 \ge Q_1^{L,\nu,0} = \frac{\mu^2e^{-\mu}}{\mu\nu-\nu^2}(Q_\nu
e^{\nu}-Q_\mu
e^\mu\frac{\nu^2}{\mu^2}-\frac{\mu^2-\nu^2}{\mu^2}Y_0).
\end{aligned}
\end{equation}

We remark that Eq.~\eqref{Decoy:VWY1Bound} can be used to provide a
simple derivation of the fraction of ``tagged photons'' $\Delta$
found in Wang's paper \cite{WangDecoy},
\begin{equation}\label{Decoy:FromQ1ToDel}
\begin{aligned}
\Delta &= \frac{Q_\nu-Y_0e^{-\nu}-Y_1\nu e^{-\nu}}{Q_\nu} \\
       &\le \frac{Q_\nu-Y_0e^{-\nu}-Y_1^{L,\nu,0}\nu e^{-\nu}}{Q_\nu} \\
       &= \frac{Q_\nu-Y_0e^{-\nu}-\frac{\mu e^{-\nu}}{\mu-\nu}(Q_\nu e^{\nu}-Q_\mu e^{\mu}\frac{\nu^2}{\mu^2}-\frac{\mu^2-\nu^2}{\mu^2}Y_0)}{Q_\nu} \\
%       &= 1-[\frac{\mu^2e^{-\mu}}{\mu\nu-\nu^2}(Q_\nu e^{\nu}-Q_\mu e^\mu\frac{\nu^2}{\mu^2})-\frac{\mu}{\nu}Y_0e^{-\nu}]/{Q_\nu} \\
%       &= \frac{Q_\nu-\frac{\mu e^{-\nu}}{\mu-\nu}(Q_\nu e^{\nu}-Q_\mu e^{\mu}\frac{\nu^2}{\mu^2})}{Q_\nu}+\frac{\mu e^{-\nu}Y_0}{\nu Q_\mu} \\
%       &= \frac{\mu}{\mu'-\mu}(\frac{\mu e^{-\mu}Q_{\mu'}}{\mu'e^{-\mu'}Q_\mu}-1)+\frac{\mu e^{-\nu}Y_0}{\nu Q_\mu} \\
       &= \frac{\nu}{\mu-\nu}(\frac{\nu e^{-\nu}Q_{\mu}}{\mu e^{-\mu}Q_\nu}-1)+\frac{\nu e^{-\nu}Y_0}{\mu
       Q_\nu}.
\end{aligned}
\end{equation}
Indeed, if we replace $\nu$ by $\mu$ and $\mu$
by $\mu'$, Eq.~\eqref{Decoy:FromQ1ToDel}
will be exactly the same as Eq.~\eqref{Intro:WangDelta}.

According to Eq.~\eqref{Decoy:e1Bound}, the upper bound of $e_1$ is
given by
\begin{equation}\label{Decoy:VWe1Bound}
\begin{aligned}
e_1 \le e_1^{U,\nu,0} = \frac{E_\nu Q_\nu
e^{\nu}-e_0Y_0}{Y_1^{L,\nu,0}\nu}.
\end{aligned}
\end{equation}

%We remark that Eqs.~\eqref{Decoy:VWY1Bound} and
%\eqref{Decoy:VWe1Bound} may not be optimal to lower bound the key
%generation rate.  More general discussion is given in
%Appendix~\ref{BetterVW} and a better bound is given in
%Appendix~\ref{VWTighterP}.

\textbf{Deviation from theoretical values:} Considering the
approximation $\eta\ll1$ and taking the first order in $\nu$,
similar to Eqs.~\eqref{Decoy:Y1DevAppro} and
\eqref{Decoy:e1DevAppro}, the theoretical deviations of Vacuum+Weak
decoy method are given by,
$$
%\begin{equation}\label{Decoy:Y1e1DevAppro}
\begin{aligned}
Y_1\beta_{Y1}
&= Y_1^{L,0}-Y_1^{L,\nu,0}\\
&= Y_0+\eta-\frac{\mu}{\mu\nu-\nu^2}(Q_\nu
e^{\nu}-Q_\mu e^\mu\frac{\nu^2}{\mu^2}-\frac{\mu^2-\nu^2}{\mu^2}Y_0)\\
%&\approx 1-\frac{\mu}{\mu\nu-\nu^2}(\nu e^{\nu}-\mu
%e^\mu\frac{\nu^2}{\mu^2})\\
%&\approx \frac\nu{\mu-\nu}(e^\mu-1-\mu)\cdot\eta +
%\frac\nu{\mu(\mu-\nu)}(e^\mu-1-\mu-\frac{\mu^2}2)\cdot Y_0. \\
&\approx \frac\nu{\mu}(e^\mu-1-\mu)\cdot\eta +
\frac\nu{\mu^2}(e^\mu-1-\mu-\frac{\mu^2}2)\cdot Y_0 \\
%&\approx (e^\mu-1-\mu)\frac\nu{\mu} - \frac{Y_0}{2Y_1}\nu. \\
e_1\beta_{e1} &= e_1^{U,\nu,0}-e_1^{U,0}\\
%&= \frac{E_\nu Q_\nu e^{\nu}-e_0Y_0}{(1-\beta_{Y1})Y_1\nu}-e_1\\
%&\approx \frac{e_1Y_1\nu+e_2Y_2\nu^2/2}{e_1Y_1\nu}+\beta_{Y1}-1 \\
%&\approx \frac{e_2Y_2\nu^2/2}{e_1Y_1\nu}+\beta_{Y1} \\
%&\approx (e^\mu-1)\frac\nu\mu-\frac{Y_0}{2Y_1}\nu. \\
%&\approx \frac\nu{\mu}(e^\mu-1-\mu)\cdot\eta +
%\frac\nu{\mu^2}(e^\mu-1-\mu-\frac{\mu^2}2)\cdot Y_0. \\
&\approx e_1\beta_{Y1}+\nu(e_1-\frac{e_0Y_0}{2Y_1}), \\
\end{aligned}
%\end{equation}
$$
from which we can see that decreasing $\nu$ will improve the
estimation of $Y_1$ and $e_1$. So, the smaller $\nu$ is, the higher
the key generation rate $R$ is. Later in section \ref{Stab}, we will
take into account of statistical fluctuations and give an estimation
on the optimal value of $\nu$ which maximizes the key generation
rate.

\subsection{One decoy state} \label{OneDecoy}
Here we will discuss a decoy state protocol with only one decoy state.
Such a protocol is easy to implement in experiments, but
may generally not be optimal.
As noted earlier, we
have successfully performed an
experimental implementation of one-decoy-state QKD in \cite{experimentaldecoy}.

\textbf{A simple proposal:} A simple method to analyze one decoy
state QKd is by substituting an upper bound of $Y_0$ into
Eq.~\eqref{Decoy:VWY1Bound} and a lower bound of $Y_0$ into
Eq.~\eqref{Decoy:VWe1Bound} to lower bound $Y_1$ and upper bound
$e_1$.

%Alternatively, a simpler but perhaps less tight method is for Alice
%and Bob to use the signal state to estimate $e_1$. It goes as
%follows:

An upper bound of $Y_0$ can be derived from Eq.~\eqref{Model:QBER},
\begin{equation}\label{Decoy:Y0Up}
\begin{aligned}
Y_0 \le \frac{E_\mu Q_\mu e^{\mu}}{e_0}.\\
\end{aligned}
\end{equation}
Substituting the above upper bound into Eq.~\eqref{Decoy:VWY1Bound},
we get a lower bound on $Y_1$

\begin{equation}\label{Decoy:OneY1e1Boundmu}
Y_1 \ge \bar{Y}_1^{L, \nu} = \frac{\mu}{\mu\nu-\nu^2}(Q_\nu e^{\nu}-Q_\mu
e^\mu\frac{\nu^2}{\mu^2}-E_\mu Q_\mu
e^{\mu}\frac{\mu^2-\nu^2}{e_0\mu^2}).
\end{equation}

A simple lower bound on $e_1$ can be derived as follows:
\begin{equation}\label{Decoy:Onee1Boundmu}
e_1 \le \bar{e}_1^{U,\nu} = \frac{E_\mu Q_\mu e^{\mu}}{Y_1^{L,\mu,0}\mu}.
\end{equation}

Now, by substituting Eqs.~\eqref{Decoy:OneY1e1Boundmu} and
\eqref{Decoy:Onee1Boundmu} into Eq.~\eqref{Intro:KeyRate}, one
obtains a simple lower bound of the key generation rate. The above
lower bound has recently been used in our experimental decoy state
QKD paper \cite{experimentaldecoy}.
[In our experimental decoy QKD paper \cite{experimentaldecoy},
we simplify our notation by denoting $\bar{Y}_1^{L, \nu}$
by simply $Y_1^L$ and $\bar{e}_1^{U,\nu} $ by $e_1^U$.]

%\textbf{Method~One: Using weak decoy state to estimate $e_1$:}
\textbf{Tighter bound:} Another method is to apply the results of
Vacuum+Weak decoy described in Subsection~\ref{V+W}.

Let's assume that Alice and Bob perform Vacuum+Weak decoy method,
but they prepare very few states as the vacuum state. So they cannot
estimate $Y_0$ very well. We claim that a single decoy protocol is
the same as a Vacuum+Weak decoy protocol, except that we do not know
the value of $Y_0$. Since Alice and Bob do not know $Y_0$, Eve can
pick $Y_0$ as she wishes. We argue that, on physical ground, it is
advantageous for Eve to pick $Y_0$ to be zero. This is because Eve
may gather more information on the single-photon signal than the
vacuum. Therefore, the bound for the case $Y_0 =0$ should still
apply to our one-decoy protocol. [We have explicitly checked
mathematically that our following conclusion is correct, after lower
bounding Eq.~\eqref{Decoy:Priv} directly.]
%in %Appendix~\ref{LowerY0}
For this reason, Alice and Bob can derive a bound on the key
generation rate, $R$, by substituting the following values of
$Y_1^{trial}$ and $e_1^{trial}$ into Eq.~\eqref{Intro:KeyRate}.

\begin{equation}\label{Decoy:OneY1e1}
\begin{aligned}
Y_1^{trial} &= \frac{\mu}{\mu\nu-\nu^2}(Q_\nu e^{\nu}-Q_\mu
e^\mu\frac{\nu^2}{\mu^2}) \\
e_1^{trial} &= \frac{E_\nu Q_\nu e^{\nu}}{Y_1^{trial}\nu}.
%e_1 \le e_1^{U,\nu} &=
%\min\{\frac{E_\mu Q_\mu e^{\mu}}{Y_1^{L,\mu,0}\mu},\frac{E_\nu Q_\nu
%e^{\nu}}{Y_1^{L,\nu,0}\nu}\}.
\end{aligned}
\end{equation}
%Here we slightly modify Eq.~\eqref{Decoy:OneY1e1} because Alice
%and Bob can use either signal state or decoy state to estimate
%$e_1$.

\subsection{Example} %\label{Example}
Let us return to the two-decoy-state protocol.
In Eqs.~\eqref{Decoy:Y1BoundAppro} and \eqref{Decoy:e1BoundAppro},
we have showed that two-decoy-state method is optimal in the
asymptotic case where $\nu_1, \nu_2 \rightarrow 0$, in the sense
that its key generation rate approaches the most general decoy state method
of having infinite number of decoy states. Here, we will
give an example to show that, even in the case of finite $\nu_1$ and
$\nu_2$, the performance of our two-decoy-state method is only
slightly worse than the perfect decoy method. We will use the model
in section \ref{Model} to calculate the deviations of the estimated
values of $Y_1$ and $e_1$ from our two-decoy-state method from the
correct values. We use the data of GYS \cite{GYS} with key
parameters listed in Table \ref{OptMu:Tab:Data}.

\begin{figure}[hbt]
\centering \resizebox{12cm}{!}{\includegraphics{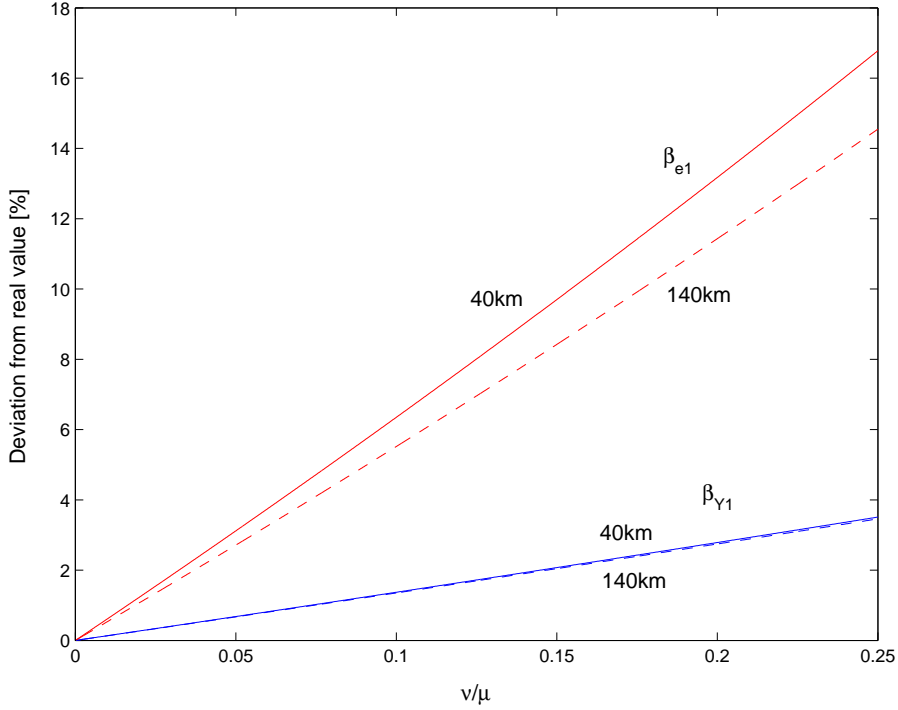}}
\caption{(Color Online) The solid lines show the relative deviations
of $Y_1^{L,\nu_1,\nu_2}$ and $e_1^{U,\nu_1,\nu_2}$ from the
asymptotic values (i.e., the case $\nu_1, \nu_2 \to 0$) as functions
of $\nu/ \mu$ (where $\nu = \nu_1$) with the fiber length 40km and
the dashed lines show the case of 140km. The bounds
$Y_1^{L,\nu_1,\nu_2}$ and $e_1^{U,\nu_1,\nu_2}$ are given by
Eqs.~\eqref{Decoy:VWY1Bound} and \eqref{Decoy:VWe1Bound}, and the
true values are given by Eqs.~\eqref{Model:Yi} and \eqref{Model:ei}.
We consider Vacuum+Weak protocol here ($\nu_1=\nu$ and $\nu_2=0$).
The expected photon number is $\mu=0.48$ as calculated from
Eq.~\eqref{OptMu:Final}. The parameters used are from GYS~\cite{GYS}
as listed in Table~\ref{OptMu:Tab:Data}.} \label{Decoy:Fig:Devnu}
\end{figure}

For simplicity, we will use a special two-decoy-state method:
Vacuum+Weak. According to Eq.~\eqref{OptMu:Final}, the optimal
expected photon number is $\mu=0.48$. We change the expected photon
number of weak decoy $\nu$ to see how the estimates, described by
Eqs.~\eqref{Decoy:VWY1Bound} and \eqref{Decoy:VWe1Bound}, deviate
from the asymptotic values, Eqs.~\eqref{Model:Yi} and
\eqref{Model:ei}. The deviations are calculated by
Eqs.~\eqref{Decoy:Y1Dev} and \eqref{Decoy:e1Dev}. The results are
shown in Figure \ref{Decoy:Fig:Devnu}. From Figure \ref{Decoy:Fig:Devnu},
we can see that the
estimate for $Y_1$ is very good. Even at $\nu/\mu=25\%$, the
deviation is only $3.5\%$. The estimate for $e_1$ is slightly
worse. The deviation will go to $16.8\%$ when $\nu/\mu=25\%$. The
deviations do not change much with fiber length. Later in Section~4,
we will discuss how to choose optimal $\nu$ when statistical
fluctuations due to a finite experimental time are taken into
account.

\begin{figure}[hbt]
\centering \resizebox{12cm}{!}{\includegraphics{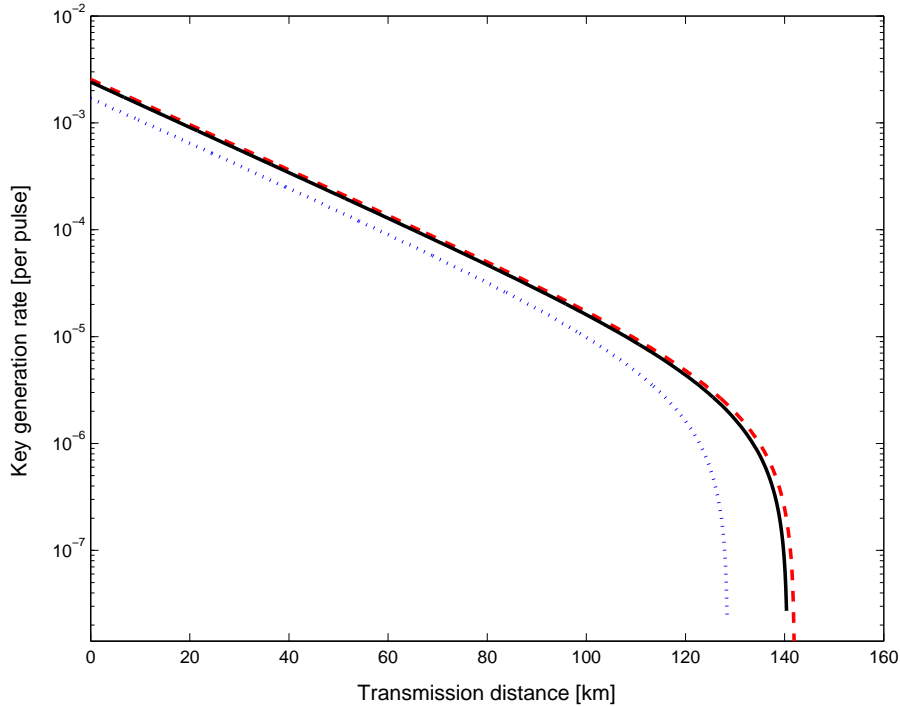}}
\caption{(Color Online) The dashed line shows the asymptotic decoy
state method (with infinite number of decoy states) with a maximal
secure distance of $142.05km$, using Eq.~\eqref{Intro:KeyRate}. The
solid line shows our Vacuum+Weak decoy method,
Eq.~\eqref{Decoy:RateLVW}, with $\mu=0.48$, $\nu_1=0.05$ and
$\nu_2=0$. It uses a strong version of GLLP and its maximal distance
is $140.55km$. The dotted line shows the asymptotic case of Wang's
decoy method, Eq.~\eqref{Decoy:WangRate} with $\mu=0.30$. It uses a
weak version of GLLP and its maximal distance is about $128.55km$.
This shows that our Vacuum+Weak decoy protocol performs very close
to the asymptotic limit and performs better than even the asymptotic
case of Wang's decoy method. The data are from GYS~\cite{GYS} as
listed in Table~\ref{OptMu:Tab:Data}.} \label{Decoy:Fig:GYS}
\end{figure}

Let $R^L$ denote for the lower bound of key generation rate,
according to \eqref{Intro:KeyRate},
\begin{equation}\label{Decoy:RateLVW}
\begin{aligned}
R^L = q\{
-Q_{\mu}f(E_{\mu})H_2(E_{\mu})+Q_1^{L,\nu,0}[1-H_2(e_1^{U,\nu,0})]\},
\end{aligned}
\end{equation}
where $q=\frac12$ with standard BB84. The parameters can be
calculated from Eqs.~\eqref{Model:Gain}, \eqref{Model:QBER},
\eqref{Decoy:VWQ1Bound} and \eqref{Decoy:VWe1Bound} and use
$f(e)=1.22$, which is the upper bound of $f(e)$ in secure distance
for this experiment \cite{BS}. Eq.~\eqref{Model:Eta} shows the
relationship between $\eta$ and distance. The results are shown in
Figure \ref{Decoy:Fig:GYS}.

Now, from Figure~\ref{Decoy:Fig:GYS}, we can see that even with finite
$\nu$ (say, $0.05$), Vacuum+Weak protocol performs very close to the
asymptotic one.

We note that Wang \cite{WangDecoy} has also studied a decoy state
protocol, first proposed
by us \cite{Decoy}, with only two decoy states for the special case where one
of them is a vacuum. In \cite{WangDecoy} the second decoy state is
used to estimate the multi photon fraction $\Delta$ and use the
formula directly from GLLP \cite{GLLP} to calculate the key
generation rate by Eq.~\eqref{Intro:KeyRateWang}.

In Figure \ref{Decoy:Fig:GYS}, we compare the key generation rates
of our two-decoy-state method and Wang's method \cite{WangDecoy} and
find that our method performs better. In what follows, we compare
the differences between our method and that of Wang.

%Later in Figure \ref{Stab:Fig:DevRdis},
%we will show that even if taking the statistical fluctuation into
%account, the weak decoy state has the big advantage of the
%asymptotic case of strong decoy method through the whole distance
%range.

\begin{itemize}
    \item We consider error correction inefficiency
$f(e)$ for practical protocols. Wang did not consider this real-life
issue. For a fair comparison, we add this factor to
Eq.~\eqref{Intro:KeyRateWang}
\begin{equation}\label{Decoy:WangRate}
\begin{aligned}
R \geq q Q_{\mu}
\{-f(E_\mu)H_2(E_{\mu})+(1-\Delta)[1-H_2(\frac{E_\mu}{1-\Delta})]\}.
\end{aligned}
\end{equation}
    \item Apparently, the value of $\mu$ was chosen in \cite{WangDecoy} in an
ad hoc manner, whereas we performed optimization in Subsection 3.1
and found that for GYS, the optimal value of $\mu=0.48$ for our
two-decoy-state method. Now, the best
(asymptotic) estimate Wang's method can make is that $\Delta=\mu$
when $\mu'\rightarrow\mu$. For a fair comparison, we have performed
an optimization of Wang's
asymptotic result Eq.~\eqref{Decoy:WangRate} as well (similar to
Subsection~\ref{OptMu}) and found that
the value $\mu \approx 0.30$ optimizes the key generation rate in Wang's method.
    \item In Eqs.~\eqref{Decoy:Y1BoundAppro} and \eqref{Decoy:e1BoundAppro},
we show that our two-decoy-state method approaches a fundamental
limit of the decoy state (the infinite decoy state protocol)
while the asymptotic result in Wang \cite{WangDecoy} is strictly
bounded away from the fundamental limit.
Even with a finite $v_1$, our Vacuum+Weak
protocol is better than Wang's asymptotic case.
    \item Why do we get a stronger result than Wang's \cite{WangDecoy}?
Wang did not estimate $e_1$ and used $E_{\mu}/(1-\Delta)$
as the upper bound of $e_1$ (This corresponds to
a weak version of GLLP \cite{GLLP}). We estimate $e_1$
more accurately following GLLP (a strong version of GLLP result).

\end{itemize}

%%%%%%%%%%%%%%%%%%%%%%%%%%%%%%%%%%%%%%%%%%%%%%%%%%%%%%%%%%%%%%%%%%%%%%%
% Stability
%%%%%%%%%%%%%%%%%%%%%%%%%%%%%%%%%%%%%%%%%%%%%%%%%%%%%%%%%%%%%%%%%%%%%%%
\section{Statistical Fluctuations} \label{Stab}
In this section, we would like to discuss the effect of finite data
size in real life experiments on our estimation process for $Y_1$
and $e_1$.
We will also discuss how statistical fluctuations
might affect our choice of
$\nu_1$ and $\nu_2$.
We will provide a list of those fluctuations and discuss how we will
deal with them.
We remark that Wang \cite{WangDecoy} has previously considered the issue of
fluctuations of $Y_1$.

All real-life experiments are done in a finite time. Ideally, we
would like to consider a QKD experiment that can be performed within
say a few hours or so. This means that our data size is finite.
Here, we will see that this type of statistical fluctuations is a
rather complex problem. We do not have a full solution to the
problem. Nonetheless, we will provide some rough estimation based on
standard error analysis which suggests that the statistical
fluctuation problem of the two-decoy-state method for a QKD
experiment appears to be under control, if we run an experiment over
only a few hours.

\subsection{What parameters are fluctuating?}
Recall that from Eq.~\eqref{Intro:KeyRate}, there are four
parameters that we need to take into account: the gain $Q_\mu$ and QBER
$E_\mu$ of signal state and the gain $Q_1$ and QBER $e_1$
of single photon sate. The gain of signal state $Q_\mu$ is measured directly
from experiment. We note that the fluctuations of
the signal error
rate $E_{\mu}$ is not important because $E_{\mu}$ is {\it not} used
at all in the
estimation of $Y_1$ and $e_1$. (See Eqs.~\eqref{Decoy:Y1Bound}
and \eqref{Decoy:e1Bound} or Eqs.~\eqref{Decoy:VWQ1Bound}
and \eqref{Decoy:VWe1Bound}.)
Therefore, the important issue is the
statistical fluctuations of $Q_1$ and $e_1$ due to the finite data
size of signal states and decoy states.

To show the complexity of the problem, we will now discuss the
following five sources of fluctuations. The first thing to notice is
that, in practice, the intensity of the lasers used by Alice will be
fluctuating. In other words, even the parameters $\mu$, $\nu_1$ and
$\nu_2$ suffer from small statistical fluctuations. Without hard
experimental data, it is difficult to pinpoint the extent of their
fluctuations. To simplify our analysis, we will ignore their
fluctuations in this paper.

The second thing to notice is that so far in our analysis we have
assumed that the proportion of photon number eigenstates in each
type of state is fixed. For instance, if $N$ signal states of
intensity $\mu$ are emitted, we assume that exactly $ N \mu e^{-
\mu}$ out of the $N$ signal states are single photons. In real-life,
the number $\mu e^{- \mu}$ is only a probability, the actual number
of single photon signals will fluctuate statistically. The
fluctuation here is dictated by the law of large number though. So,
this problem should be solvable. For simplicity, we will neglect
this source of fluctuations in this paper.
[It was subsequently pointed out to us by Gottesman and Preskill that
the above two sources of fluctuations can be combined into the fluctuations
in the photon number frequency distribution of the underlying signal and
decoy states.
These fluctuations will generally average out to zero in the limit of
a large number of signals, provided that there is no systematic error
in the experimental set-up.]

The third thing to notice is, as noted by Wang \cite{WangDecoy}, the
yield $Y_i$ may fluctuate in the sense that $Y_i$ for the signal
state might be slightly different from $Y'_i$ of the decoy state. We
remark that if one uses the vacuum state as one of the decoy states,
then by observing the yield of the vacuum decoy state, conceptually,
one has a very good handle on the yield of the vacuum component of
the signal state (in terms of hypergeometric functions). Note,
however, that the background rate is generally rather low (typically
$10^{-5}$). So, to obtain a reasonable estimation on the background
rate, a rather large number (say $10^7$) of vacuum decoy states will
be needed. [As noted in \cite{Decoy}, even a $20\%$ fluctuations in
the background will have small effect on the key generation rates
and distances.] Note that, with the exception of the case $n=0$ (the
vacuum case), neither $Y_i$ and $Y_i'$ are directly observable in an
experiment. In a real experiment, one can measure only some {\it
averaged} properties. For instance, the yield $Q_{\mu}$ of the
signal state, which can be experimentally measured, has its origin
as the weighted averaged yields of the various photon number
eigenstates $Y_i$'s whereas that for the decoy state is given by the
weighted averaged of $Y'_i$'s. How to relate the observed averaged
properties, e.g., $Q_{\mu}$, to the underlying values of $Y_i$'s is
challenging question. In summary, owing to the fluctuations of $Y_i$
for $n >0$, it is not clear to us how to derive a closed form
solution to the problem.

Fourth, we note that the error rates, $e_i$'s, for the signal
can also be different from the error rates $e_i$'s for the
decoy state, due to underlying statistical fluctuations.
Actually, the fluctuation of $e_1$ appears to the dominant source
of errors in the estimation process. (See, for example, Table
\ref{Stab:Tab:Exp}.)
This is because the parameter $e_1$
is rather small (say a few percent) and it appears in combination with
another small parameter $Y_1$ in Eq.~\eqref{Model:QBER} for QBER.

Fifth, we noted that for security in the GLLP \cite{GLLP} formula (Eq.~(1)),
we need to correct phase
errors, rather than bit-flip errors. From Shor-Preskill's proof \cite{ShorPreskill},
we know that the bit-flip error rate
and the phase error rate are supposed to be the same only in the asymptotic limit.
Therefore, for a finite data set, one
has to consider statistical fluctuations.
This problem is well studied \cite{ShorPreskill}.
Since the number of signal states is generally very big, we
will ignore this fluctuation from now on.

Qualitatively, the yields of the signal and decoy states
tend to decrease exponentially with distance.
Therefore, statistical fluctuations tend to become more
and more important as the distance of QKD increases.
In general, as the distance of QKD increases,
larger and large data sizes will be needed for the reliable
estimation of $Y_1$ and $e_1$ (and hence $R$), thus
requiring a longer QKD experiment.

In this paper, we will neglect the fluctuations due to the
first two and the fifth sources listed above.
Even though we cannot find any closed form solution for the third
and fourth sources of fluctuations, it should be possible to tackle
the problem by simulations. Here, we are contented with a
more elementary analysis. We will
simply apply standard
error analysis to perform a rough estimation on the effects of
fluctuations due to the third and fourth sources. We remark that the
origin of the problem is strictly classical statistical
fluctuations. There is nothing quantum in this statistical analysis.
While standard error analysis (using essentially normal
distributions) may not give a completely correct answer, we expect
that it is correct at least in the order of magnitude.

Our estimation, which will be presented below, shows that, for
long-distance ($>100km$) QKD with our two-decoy-state protocol, the
statistical fluctuations effect (from the third and fourth sources
only) appears to be manageable. This is so provided that a QKD
experiment is run for a reasonable period of time of only a few
hours. Our analysis supports the viewpoint that our
two-decoy-state protocol is practical for real-life implementations.

We remark on passing that the actual classical memory space requirement
for Alice and Bob is rather modest ($<1 GBytes$) because at long
distance, only a small fraction of the signals will give rise to
detection events.

We emphasize that we have not fully solved the statistical
fluctuation problem for decoy state QKD. This problem
turns out to be quite
complex. We remark that this statistical
fluctuation problem will affect all earlier
results including \cite{HwangDecoy,Decoy,WangDecoy}.
In future investigations, it will be interesting to
study the issues of classical statistical fluctuations in more detail.

\subsection{Standard Error Analysis}

In what follows, we present a general procedure
for studying the statistical
fluctuations (due to the third and fourth sources noted above) by using
standard error analysis.

Denote the number of pulses (sent by Alice) for signal as $N_S$,
and for two decoy states as $N_1$ and
$N_2$. Then, the total number of pulses sent by Alice is given by
\begin{equation}\label{Stab:NTotal}
\begin{aligned}
N &= N_S+N_1+N_2.\\
\end{aligned}
\end{equation}
Then the parameter $q$ in Eq.~\eqref{Intro:KeyRate} is given by
\begin{equation}\label{Stab:q}
\begin{aligned}
q &= \frac{N_S}{2N}. \\
\end{aligned}
\end{equation}
Here we assume Alice and Bob perform standard BB84. So, there is
a factor of $\frac12$.

In practice, since $N$ is finite, the statistical
fluctuations of $Q_1$ and $e_1$ cannot be neglected. All
these additional deviations will be related to data sizes $N_S$,
$N_1$ and $N_2$ and can, in principle, be obtained from statistic analysis.
A natural question to ask is the following. Given total data size $N=const$, how
to distribute it to $N_S$, $N_1$ and $N_2$ to maximize the
key generation rate $R$? This question also relates to another one:
how to choose optimal weak decoy $\nu_1$ and $\nu_2$ to minimize
the effects of statistical fluctuations?

In principle, our optimization procedure should go as follows.
First, (this is the hard part) one needs to derive a lower bound of
$Q_1$ and an upper bound of $e_1$ (as functions of data
size $N_S$, $N_1$, $N_2$, $\nu_1$ and $\nu_2$), taking into
full account of statistical fluctuations. Second, one substitutes those
bounds to Eq.~\eqref{Intro:KeyRate} to calculate the lower bound of
the key generation rate, denoted by $R^L$.
%The lower bound of $Q_\mu$ is a function of $N_S+N_W$, the lower bound
%of $Q_1$ is a function of $N_0$, $N_W$ and $\nu$, the upper bound of
%$E_\mu$ is a function of $N_E$ and the upper bound of $e_1$ is a
%function of $N_0$, $N_W$ and $\nu$.
%The lower bounds of $Q_\mu$, $Q_1$ and the upper bounds of $E_\mu$,
%$e_1$ are functions of $N_E$, $N_0$, $N_W$ and $\nu$.
Thus, $R^L$ is a function of $N_S$, $N_1$, $N_2$, $\nu_1$ and
$\nu_2$, and will be maximized when the optimal distribution
satisfies
\begin{equation} \label{Stab:RDevN}
\begin{aligned}
\frac{\partial R^L}{\partial N_S}
 = \frac{\partial R^L}{\partial N_1} = \frac{\partial
R^L}{\partial N_2}% = \frac{\partial R^L}{\partial \nu_1} =
%\frac{\partial R^L}{\partial \nu_2}
= 0,
\end{aligned}
\end{equation}
%where $\beta_{E\mu}$, $\beta_{Y1}$ and $\beta_{e1}$ is the percent
%deviations of $E_\mu$, $Y_1$ and $e_1$, i.e., the lower bound of
%$Y_1$ is given by $Y_1(1-\beta_{Y1})$, the upper bounds of $E_\mu$
%and $e_1$ are given by $E_\mu(1+\beta_{E\mu})$ and
%$e_1(1+\beta_{e1})$. %Through statistical fluctuation analysis on
%each variable, see in Appendix, we can solve Eq.~\eqref{Stab:RDevN}
%to get optimal distribution of $N_S$, $N_E$, $N_0$ and $N_W$.
%where lower bound of $Q_\mu$ will be a function of $N_S+N_E$, upper
%bound of $E_\mu$ relates to $N_E$,
given $N=N_S+N_1+N_2=const$.

\subsection{Choice of $\nu_1$ and $\nu_2$}
Now, from the theoretical deviations of $Y_1$ and $e_1$,
Eqs.~\eqref{Decoy:Y1Dev} and \eqref{Decoy:e1Dev}, reducing $\nu$ may
decrease the theoretical deviations. We need to take statistical
fluctuations into account. Given a fixed $N_1+N_2$, reducing $\nu_1$
and $\nu_2$ will decrease the number of detection events of decoy
states, which in turns causes a larger statistical fluctuation.
Thus, there exists an optimal choice of $\nu_1$ and $\nu_2$ which
maximizes the lower bound of the key generation rate $R^L$,
$$
\frac{\partial R^L}{\partial \nu_1} =
\frac{\partial R^L}{\partial \nu_2} = 0, \\
$$
which can be simplified to
\begin{equation}\label{Stab:Optnu}
\begin{aligned}
\frac{\partial}{\partial\nu_1}\{\hat{Y}_1^{L,\nu_1,\nu_2}
[1-H_2(\hat{e}_1^{U,\nu_1,\nu_2})]\} &=0 \\
\frac{\partial}{\partial\nu_2}\{\hat{Y}_1^{L,\nu_1,\nu_2}
[1-H_2(\hat{e}_1^{U,\nu_1,\nu_2})]\} &=0, \\
%-Y_1[1-H_2(e_1)]\frac{\partial\beta_{Y1}}{\partial\nu} &-
%Y_1e_1\log_2(\frac{1-e_1}{e_1})\frac{\partial\beta_{e1}}{\partial\nu}
%=0\\
%[1-H_2(e_1)]\frac{\partial\beta_{Y1}}{\partial\nu} &+
%e_1\log_2(\frac{1-e_1}{e_1})\frac{\partial\beta_{e1}}{\partial\nu}=0 \\
%[1+\log_2(1-e_1)](\frac{e^\mu-1-\mu}{\mu}+e^\nu\frac{\partial\beta_{Q\nu}}{\partial\nu})
%&+
%\log_2(\frac{1-e_1}{e_1})(\frac{Q_\nu E_\nu e^\nu}{Y_1\nu}\frac{\partial\beta_{Q\nu E\nu}}{\partial\nu}+e_1)=0 \\
%[1+\log_2(1-e_1)](\frac{e^\mu-1-\mu}{\mu} -
%e^\nu\frac{u_\alpha}{\sqrt{2\eta N_W}}\nu^{-\frac32}) &+
%\log_2(\frac{1-e_1}{e_1})(-\frac{Q_\nu E_\nu e^\nu}{Y_1\nu}\frac{u_\alpha\eta e_{detector}}{\sqrt{2N_W}}(Q_\nu E_\nu)^{-\frac32}+e_1)=0 \\
%[1+\log_2(1-e_1)]\frac{e^\mu-1-\mu}{\mu} &+
%e_1\log_2(\frac{1-e_1}{e_1}) \\
%= [1+\log_2(1-e_1)]\frac{u_\alpha e^\nu}{\sqrt{2\eta
%N_W}}\nu^{-\frac32} &+ \log_2(\frac{1-e_1}{e_1})
%\frac{Q_\nu E_\nu e^\nu}{Y_1\nu}\frac{u_\alpha\eta e_{detector}}{\sqrt{2N_W}}
%(Q_\nu E_\nu)^{-\frac32} \\
\end{aligned}
\end{equation}
%Here we take the approximation $Y_0\ll Y_1$ and $\eta\nu\ll1$.
where $\hat{Y}_1^{L,\nu_1,\nu_2}$ and $\hat{e}_1^{U,\nu_1,\nu_2}$
are lower bound to $Y_1$ and upper bound to $e_1$ when statistical
fluctuations are considered.

Given total data size in Eq.~\eqref{Stab:NTotal}, in principle, one
can solve Eqs.~\eqref{Stab:RDevN} and \eqref{Stab:Optnu} to get
$N_S$, $N_1$, $N_2$ $\nu_1$ and $\nu_2$.

\subsection{Simulation:}
In real life, solving Eqs.~\eqref{Stab:RDevN} and \eqref{Stab:Optnu}
is a complicated problem. In what follows, we will be contented
with a rough estimation procedure using standard error analysis
commonly used by experimentalists.

\textbf{Some assumptions:} In the following, we will discuss
Vacuum+Weak decoy method only.
\begin{enumerate}
    \item The signal state is used much more often than the
    two decoy states. Given the large number of signal states, it is reasonable to
    ignore the statistical fluctuations in signal states.
    \item We assume that the decoy state used in the actual
    experiment is conceptually
    only a part of an infinite population of decoy states.
    There are underlying values for $Q_{\nu}$ and $E_{\nu}$ as defined by
    the population of decoy states. In each realization, the decoy state
    allows us to obtain some estimates for these underlying $Q_{\nu}$
    and $E_{\nu}$ . Alice and Bob can use the fluctuations of $Q_{\nu}$,
$E_{\nu}$ to calculate the fluctuation of the estimates of $Y_1$
and $e_1$.
    \item We neglect the change of $f(E_{\mu})$ due to small change in
$E_\mu$.
%    \item Normally $Y_0\ll Y_1$, so we can neglect the the influence
%of $Y_0$'s fluctuation on $Y_1$.
    \item When the number of events (e.g. the total detection
    event of the vacuum decoy state) is large
(say $>50$), we assume that the statistical characteristic of a parameter can be
described by a {\it normal} distribution.
\end{enumerate}
%Notice that the second assumption is not good.

%Normal
%distribution is a good description of the detection events, when the
%counts number is large ($>50$). %We use normal distribution to
%describe the various detection events.

We will use the experiment parameters in Table \ref{OptMu:Tab:Data}
and show numerical solutions of Eqs.~\eqref{Stab:NTotal},
\eqref{Stab:RDevN} and \eqref{Stab:Optnu}. We pick the total data
size to be $N=6\times10^9$. Now, the GYS experiment \cite{GYS} has a
repetition rate of $2 MHz$ and an up time of less than $50\%$
\cite{Private}. Therefore, it should take only a few hours to
perform our proposed experiment. The optimal $\mu=0.48$ can be
calculated by Eq.~\eqref{OptMu:Final} and we use $f(e)=1.22$.

In the fiber length of %$45.44km$ ($\eta=5\times10^{-3}$) and
$103.62km$ ($\eta=3\times10^{-4}$), the optimal pulses distribution
of data, $\nu$ and the deviations from perfect decoy method
%can be calculated by Eqs.~\eqref{Stab:NTotal}, \eqref{Stab:RDevN} and
%\eqref{Stab:Optnu}, which
are listed in Table \ref{Stab:Tab:Exp}.
\begin{table}[h]\center
\begin{tabular}{|c|c|c|c|c|c|c|c|c|c|c|}
%\hline
%$l$ & $\mu$ & $u_\alpha$ & $N$ & $N_S$ & $N_E$ & $N_1$ & $N_2$ \\
%\hline
%$45.44km$ & $0.48$ & $10$ & $6\times10^{9}$ & $5.1\times10^{9}$ & $2.4\times10^{8}$ & $5.4\times10^{8}$ & $1.2\times10^{8}$ \\
%\hline
%$\eta$ & $\nu$ & $\tilde{B}[bits]$ & $\beta_{Y0}$ & $\beta_{Y1}$ & $\beta_{e1}$ & $\beta_{R}$ &  \\
%\hline
%$5\times10^{-3}$ & $0.053$ & $2.12\times10^{5}$ & $70.01\%$ & $4.66\%$ & $34.31\%$ & $42.03\%$ &  \\
%\hline
\hline
$l$ & $\mu$ & $u_\alpha$ & $N$ & $N_S$ & $N_1$ & $N_2$ \\
\hline
$103.62km$ & $0.479$ & $10$ & $6\times10^{9}$ & $3.98\times10^{9}$ & $1.76\times10^{9}$ & $2.52\times10^{8}$ \\
\hline
$\eta$ & $\nu$ & $\tilde{B}[bits]$ & $\beta_{Y0}$ & $\beta_{Y1}$ & $\beta_{e1}$ & $\beta_{R}$  \\
\hline
$3\times10^{-4}$ & $0.127$ & $2.17\times10^{4}$ & $48.31\%$ & $7.09\%$ & $97.61\%$ & $74.11\%$  \\
\hline
\end{tabular}
\caption{The pulse number distribution and $\nu$ are calculated from
Eqs.~\eqref{Stab:RDevN} and \eqref{Stab:Optnu}. $\tilde{B}$ is the
lower bound of final key bits. All results are obtained by numerical
analysis using MatLab. The variable $\tilde{\beta}_{Y1}$ denotes
the relative error in our estimation process of
$Y_1$ from its true value by using the
data from a finite experiment. This relative error
originates from statistical fluctuations.
This definition contrasts with the definition of $\beta_{Y1}$
in Eq.~\eqref{Decoy:Y1Dev} which refers to the relative difference
between the values of $Y_1$ for the case i) where $\nu_1$ and $\nu_2$ are
finite and the case ii) where $\nu_1$ and $\nu_2$ approach zero.
Similarly, other $\beta$'s denote the relative errors in
our estimates for the corresponding variables in the subscript of $\beta$.
All the statistical fluctuation is of the
confidence interval of ten standard deviations
(i.e., $1-1.5\times10^{-23}$).  The data come from
GYS~\cite{GYS}, listed in Table~\ref{OptMu:Tab:Data}.}
\label{Stab:Tab:Exp}
\end{table}

For each fiber length we can solve Eqs.~\eqref{Stab:RDevN} and
\eqref{Stab:Optnu} to get $N_S$, $N_E$, $N_1$, $N_2$ and $\nu$.

\begin{figure}[hbt]
\centering \resizebox{12cm}{!}{\includegraphics{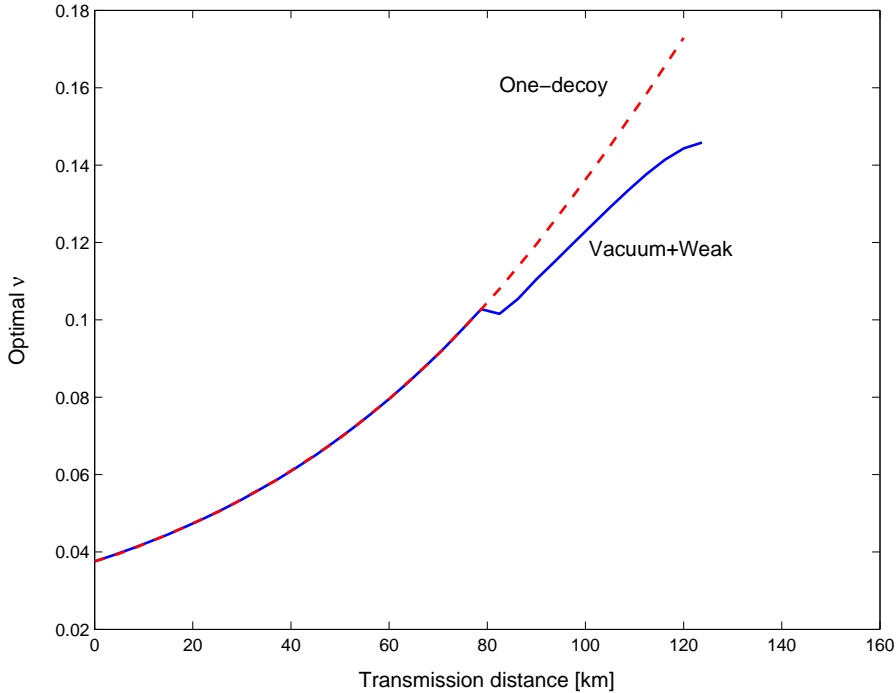}}
\caption{(Color Online) The solid line shows the simulation result
of the Vacuum+Weak protocol (Eqs.~\eqref{Decoy:VWY1Bound} and
\eqref{Decoy:VWe1Bound}) with statistical fluctuations. The dashed
line shows the result for one-decoy-state method
(Eqs.~\eqref{Decoy:OneY1e1}). Here, we pick the data size (total
number of pulses emitted by Alice) to be $N=6\times10^{9}$. We find
the optimal $\nu$'s for each fiber length by numerically solving
Eqs.~\eqref{Stab:NTotal}, \eqref{Stab:RDevN} and \eqref{Stab:Optnu}.
The confidence interval for statistical fluctuation is ten standard
deviations (i.e., $1-1.5\times10^{-23}$). The data are from
GYS~\cite{GYS} as listed in Table~\ref{OptMu:Tab:Data}. The expected
photon number of signal state is calculated by
Eq.~\eqref{OptMu:Final}, getting $\mu=0.48$. The second decoy state
(vacuum decoy) becomes useful at 82km.
%The strange behavior around 80km is due to the fact that the second
%decoy state (vacuum decoy) becomes useful there, before which one
%decoy state is sufficient to lower bound the key rate as discussed
%in Subsection~\ref{OneDecoy}.
} \label{Stab:Fig:DevNudis}
\end{figure}

Figure~\ref{Stab:Fig:DevNudis} shows how the optimal $\nu$ changes with
fiber length. We can see that the optimal $\nu$ is small ($\sim0.1$)
through the whole distance. In fact, it starts at a value $\nu\approx 0.04$
at zero distance and increases almost linearly with the distance.

\begin{figure}[hbt]
\centering \resizebox{12cm}{!}{\includegraphics{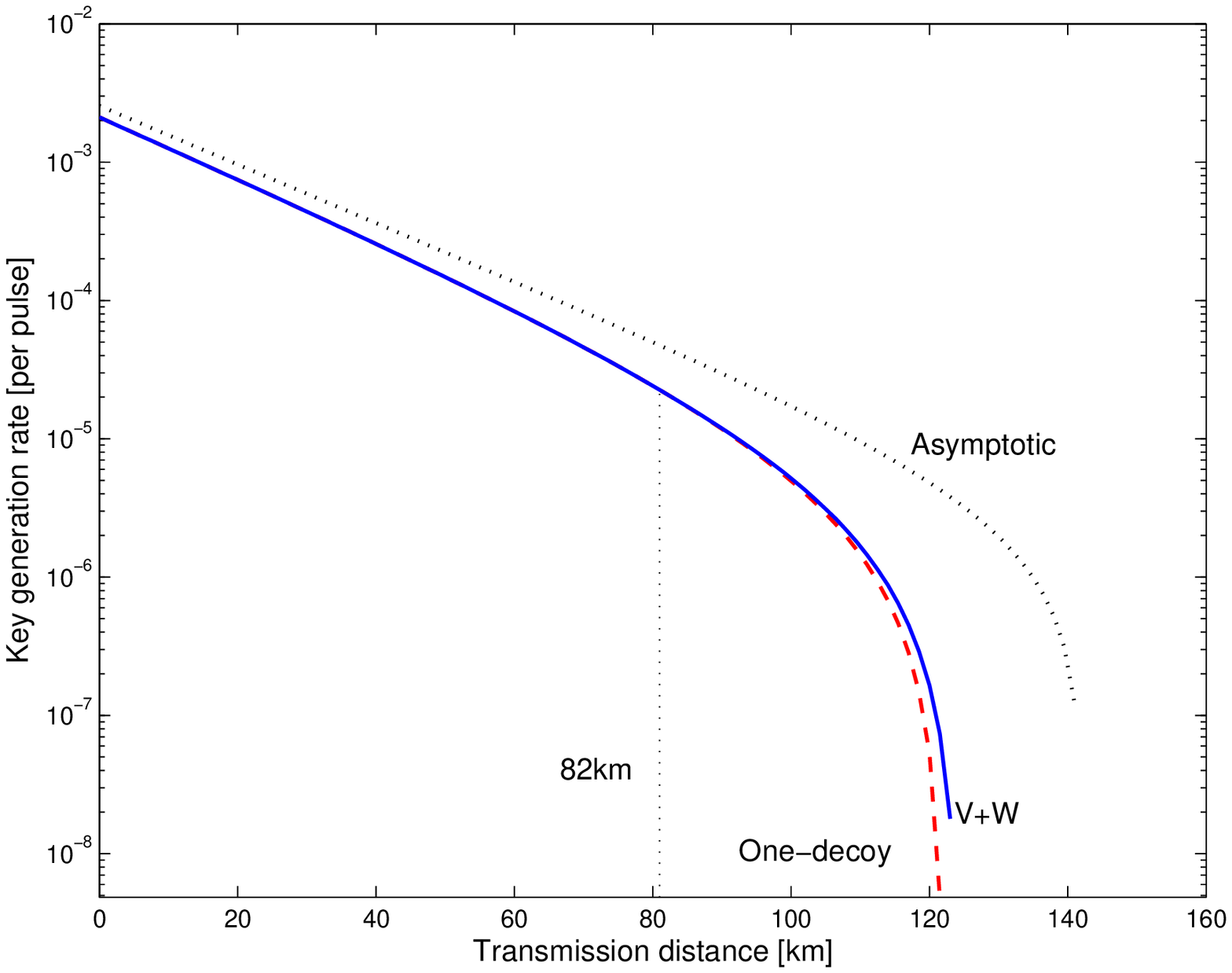}}
\caption{(Color Online) The dotted line shows the performance of
perfect decoy state method (with infinite number of decoy states and
no statistical fluctuations). The maximal distance is about 142km.
The solid line shows the simulation result of the Vacuum+Weak
protocol (Eqs.~\eqref{Decoy:VWY1Bound} and \eqref{Decoy:VWe1Bound})
with statistical fluctuations. Its maximal distance is about 125km.
The dashed line shows the result for one-decoy-state method
(Eqs.~\eqref{Decoy:OneY1e1}) with maximal distance 122km. We pick a
data size (i.e., total number of pulses emitted by Alice) to be $N=
6 \times 10^9$. Note that even with statistical fluctuations and a
rather modest data size, our Vacuum+Weak decoy protocol performs
rather close to asymptotic limit, particularly at short distances.
The second decoy state (vacuum decoy) becomes useful at 82km. The
data are from GYS~\cite{GYS} as listed in
Table~\ref{OptMu:Tab:Data}. The expected photon number of signal
state is calculated by Eq.~\eqref{OptMu:Final}, getting $\mu=0.48$.}
\label{Stab:Fig:Rdist}
\end{figure}

Figure~\ref{Stab:Fig:Rdist} shows Vacuum+Weak with statistical
fluctuations as compared to the asymptotic case of infinite
decoy state and without statistical fluctuations. We can see that
even taking into account the statistical fluctuations, the
Vacuum+Weak protocol is not far from the asymptotic result. In particular,
in the short distance region, our two-decoy-state method
with statistical fluctuations approaches
the performance of the asymptotic limit of infinite decoy
states and no statistical fluctuations. This is so
because the channel is not that lossy and statistical fluctuations
are easily under control. This fact highlights the feasibility of
our proposal.

\begin{figure}[hbt]
\centering \resizebox{12cm}{!}{\includegraphics{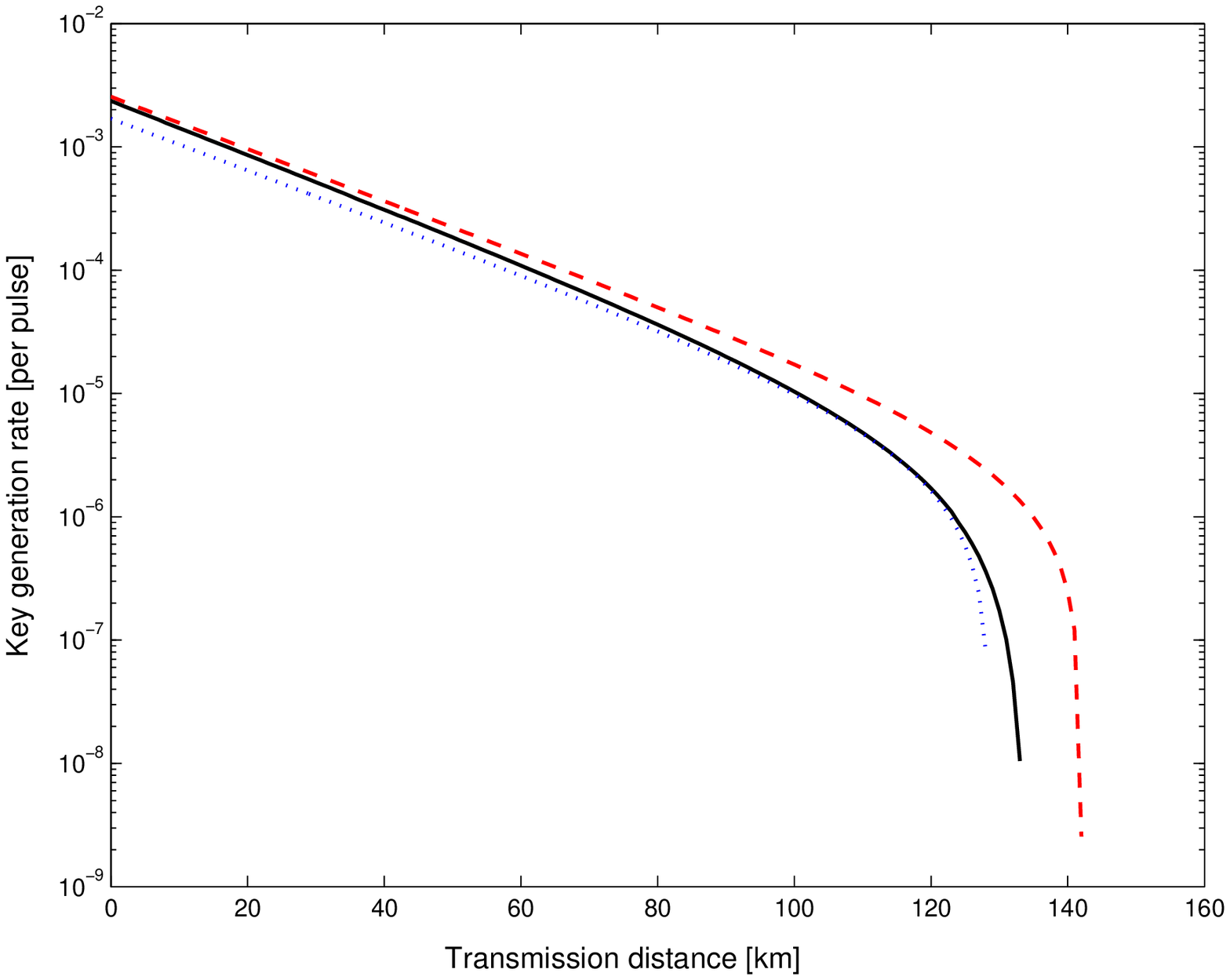}}
\caption{(Color Online) Here, we consider the data size (i.e., the
number of pulses emitted by Alice) to be $N= 8.4 \times 10^{10}$,
following Wang \cite{WangDecoy}. The dashed line shows the
performance of perfect decoy state method. Its maximal distance is
142km. The solid line shows the simulation result of the Vacuum+Weak
decoy state method with statistical fluctuations. Its maximal
distance is 132km. The dotted line shows the asymptotic case (i.e.,
an idealized version) of Wang's method. Its maximal distance is
128.55km. This figure shows clearly that with a data size $N=8.4
\times 10^{10}$, our protocol, which considers statistical
fluctuations, performs better even than the idealized version of
Wang's protocol, where statistical fluctuations are neglected. For
our asymptotic case and two-decoy with statistical fluctuation
$\mu=0.48$, and for Wang's asymptotic case $\mu=0.3$, which are
optimized.} \label{Stab:Fig:RdistComp}
\end{figure}

Wang~\cite{WangDecoy} picked the total data size
$N=8.4\times10^{10}$. For long distance QKD,
this will take more than one day of
experiment with the current
GYS set-up~\cite{GYS}. In order to perform
a fair comparison with Wang\cite{WangDecoy}'s result,
we will now the data size $N=8.4\times10^{10}$.
Figure~\ref{Stab:Fig:RdistComp} shows $R^L$ vs. fiber length $l$
with $N=8.4\times10^{10}$ fixed and compares our Vacuum+Weak protocol with
Wang's result.

\begin{figure}[hbt]
\centering \resizebox{12cm}{!}{\includegraphics{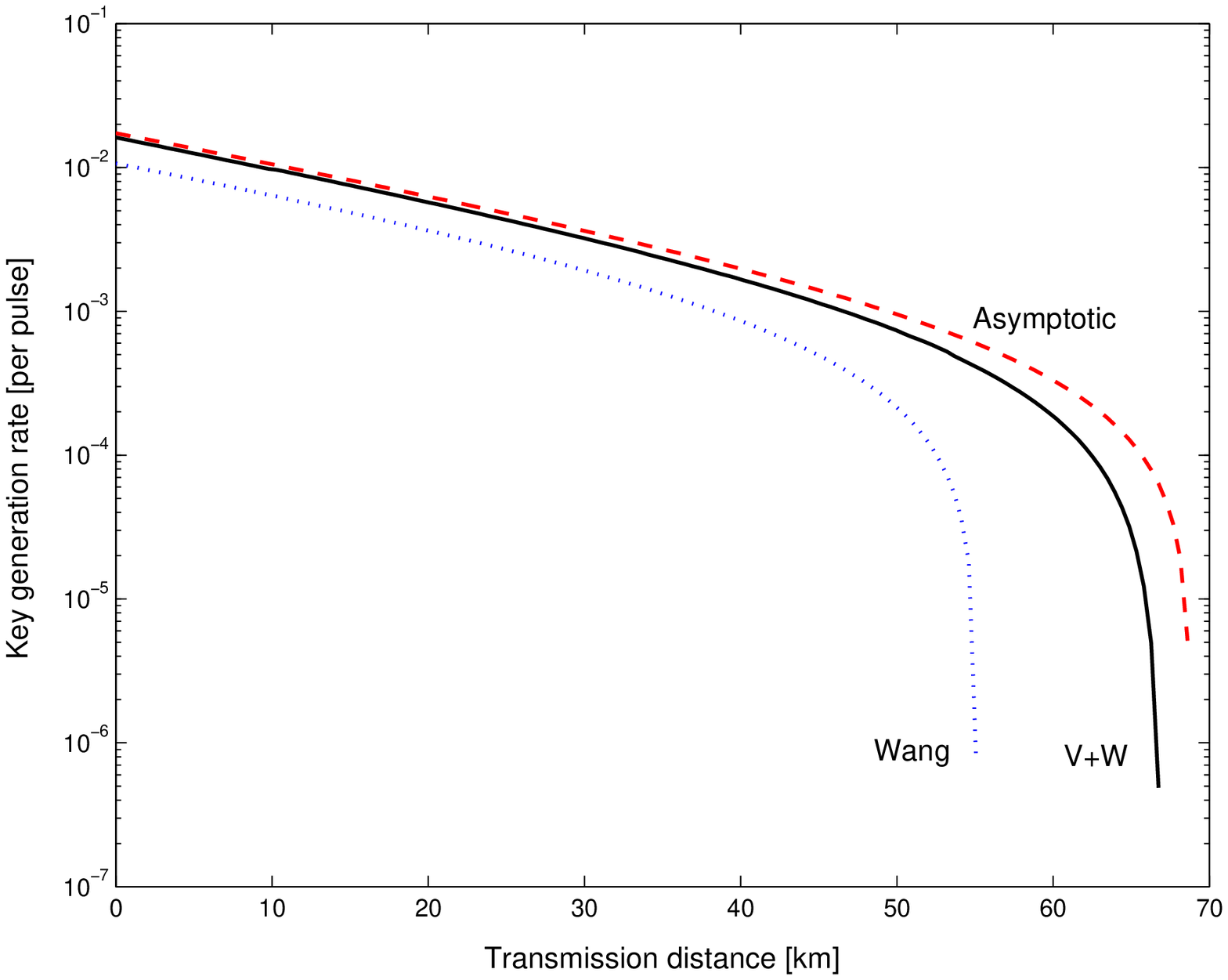}}
\caption{(Color Online) Here, we compare various protocols using the
parameters in KTH~\cite{KTH}, listed in Table~\ref{OptMu:Tab:Data}
and \cite{Lutkenhaus}. The dashed line shows the performance of
perfect decoy state method. It has a maximal secure distance of
about 68.6km. The solid line shows the simulation result of the
Vacuum+Weak decoy state method with statistical fluctuations. The
maximal distance is about 67.2km). The dotted line shows the
asymptotic case (i.e., neglecting statistical fluctuations) of
Wang's method whose maximal distance is about 55.5km.  For our
asymptotic case and two-decoy with statistical fluctuation
$\mu=0.77$, and for Wang's asymptotic case $\mu=0.43$, which are
optimized.} \label{Stab:Fig:RdistKTH}
\end{figure}

Comments:
\begin{itemize}
    \item Wang~\cite{WangDecoy} chooses the value of $\mu$ in an ad hoc manner. Here
we note that, for
Wang's asymptotic case, the optimal choice of $\mu$ is $\mu\in[0.25,0.3]$
    \item
Even if we choose $\mu\in[0.25,0.3]$, the
maximal secure distance of Wang's asymptotic case is still less than
our two-decoy-state method with statistical fluctuations.
In other words, the performance of our two-decoy-state method with
statistical fluctuations is still better than the the asymptotic
value (i.e., without considering statistical fluctuations) given by
Wang's method.
    \item Note that GYS~\cite{GYS} has a very low background rate
($Y_0=1.7\times10^{-6}$) and high $e_{detector}$. The typical values
of these two key parameters are $Y_0=10^{-5}$ and
$e_{detector}=1\%$. If the background rate is higher and
$e_{detector}$ is lower, then our results will have more advantage
over Wang's. We illustrate this fact in
Figure~\ref{Stab:Fig:RdistKTH} by using
the data from the KTH experiment \cite{KTH}.
\end{itemize}

%%%%%%%%%%%%%%%%%%%%%%%%%%%%%%%%%%%%%%%%%%%%%%%%%%%%%%%%%%%%%%%%%%%%%%%
% Conclusion
%%%%%%%%%%%%%%%%%%%%%%%%%%%%%%%%%%%%%%%%%%%%%%%%%%%%%%%%%%%%%%%%%%%%%%%
\section{Conclusion}\label{s:Conclusion}
We studied the two-decoy-state protocol where two weak decoy states
of intensities $\nu_1$ and $\nu_2$ and a signal state with intensity
$\mu $ are employed. We derived a general formula for the key
generation rate $R$ of the protocol and showed that the
asymptotically limiting case where $\nu_1$ and $\nu_2$ tend to zero
gives an optimal key generation rate which is the same as having
infinite number of decoy states. This result seems to suggest that
there is no fundamental conceptual advantage in using more than two
decoy states. Using the data from the GYS experiment \cite{GYS}, we
studied the effect of finite $\nu_1$ and $\nu_2$ on the value of the
key generation rate, $R$. In particular, we
considerd a Vacuum+Weak protocol, proposed
in \cite{Decoy} and analyzed in \cite{WangDecoy}, where $\nu_2 =0$
and showed that $R$ does not change much
even when $\nu_1 / \mu$ is as high as $25\%$. We also derived the
optimal choice of expected photon number $\mu$ of the signal state,
following our earlier work \cite{Decoy}. Finally, we considered the
issue of statistical fluctuations due to a finite data size. We remark
that statistical fluctuations have also been considered in the recent work of
Wang \cite{WangDecoy}. Here, we
listed five different sources of fluctuations. While the problem is highly
complex, we provided an estimation based on standard error analysis.
We believe that such an analysis, while not rigorous, will give at
least the correct order of magnitude estimation to the problem.
This is so because this is a classical estimation problem. There is
nothing quantum about it. That is to say there are no subtle quantum
attacks to consider. Our
estimation showed that two-decoy-state QKD appears to be highly
practical. Using data from a recent experiment \cite{GYS}, we showed
that, even for long-distance (i..e, over 100km) QKD, only a few hours of data are
sufficient for its implementation. The memory size requirement is
also rather modest ($ < 1GBytes$).
A caveat is that we have not
considered the fluctuations of the laser intensities of Alice, i.e.,
the value of $\mu$, $\nu_1$ and $\nu_2$.
This is because we do not have reliable experimental data to
perform such an investigation.
For short-distance QKD, the effects of statistical fluctuations are
suppressed because the transmittance and useful data rate are
much higher than long-distance QKD.
Finally, we noted that statistical
fluctuations will affect our choice of decoy states $\nu_1$ and
$\nu_2$ and performed an optimization for the special case
where $\nu_2 =0$.

In summary, our investigation demonstrates that a simple two decoy state protocol
with Vacuum+Weak decoy state is highly practical and can achieve
unconditional security for long-distance (over 100km) QKD, even with
only a few hours of experimental data.

As a final note, we have also studied a simple one-decoy-state protocol.
Recently, we have experimentally implemented our one-decoy-state protocol
over 15km of Telecom fibers \cite{experimentaldecoy}, thus
demonstrating the feasibility of our proposal.

\section*{Acknowledgments}
This work was financially supported in part by
Canadian NSERC, Canada Research Chairs Program, Connaught Fund,
Canadian Foundation for Innovation, Ontario Innovation Trust,
Premier's Research Excellence Award, Canadian Institute for
Photonics Innovations, and
University of Toronto start-up grant.
We thank enlightening discussions with many colleagues including, for example,
Charles Bennett, Jean Christian Boileau, Gilles Brassard, Kai Chen,
Fr\'{e}d\'{e}ric Dupuis, Daniel
Gottesman, Jim Harrington, Won-Young Hwang, Daniel Lidar,
Jeff Kimble, Stephen M. S. Lee, Debbie Leung, Norbert
L\"{u}tkenhaus, John Preskill, Aephraim Steinberg, Kiyoshi Tamaki,
Xiang-Bin Wang,
and Zhiliang Yuan.
H.-K. Lo also thanks travel support from the Isaac Newton Institute,
Cambridge, UK for its quantum information program and from the
Institute for
Quantum Information at the California Institute of Technology
through the National Science
Foundation under grant EIA-0086038.

%%%%%%%%%%%%%%%%%%%%%%%%%%%%%%%%%%%%%%%%%%%%%%%%%%%%%%%%%%%%%%%%%%%%%%%
% Appendix
\begin{appendix}

%%%%%%%%%%%%%%%%%%%%%%%%%%%%%%%%%%%%%%%%%%%%%%%%%%%%%%%%%%%%%%%%%%%%%%%
% Vacuum+Weak is optimal
%%%%%%%%%%%%%%%%%%%%%%%%%%%%%%%%%%%%%%%%%%%%%%%%%%%%%%%%%%%%%%%%%%%%%%%
\section{Appendix} \label{Optnu2}
In this appendix, we will prove that the Vacuum+Weak decoy protocol
is optimal among the two-weak-decoy protocol. We do so by proving
that, for a fixed $\nu_1 $ (which is larger than $\nu_2$),
\begin{itemize}
    \item the lower bound $Y_1^{L, \nu_1 , \nu_2}$ can be no greater than
$Y_1^{L, \nu_1 , 0}$, see Eq.\eqref{Optnu2:Y1Bound}
    \item the upper bound $e_1^{U,\nu 1, \nu 2}$ can be no less than
$e_1^{U,\nu 1, 0}$, see Eq.~\eqref{Optnu2:e1Bound}
\end{itemize}
We will consider those bounds as given in Eqs.~\eqref{Decoy:Y1Bound}
and \eqref{Decoy:e1Bound}. In what follows, we assume the conditions
given by Eq.~\eqref{Decoy:ConditionWeak},
\begin{equation}\label{Optnu2:Condition}
\begin{aligned}
%0<\eta<1\\
0\le\nu_2<\nu_1\\
\nu_1+\nu_2<\mu.
\end{aligned}
\end{equation}

\noindent\textbf{Theorem} Given $\mu$, $\nu_1$, $\eta$, $Y_0$ and
$e_{detector}$, the lower bound of $Y_1$ given in
Eq.~\eqref{Decoy:Y1Bound},
$$
\begin{aligned}
Y_1^{L,\nu_1,\nu_2} &= \frac{\mu}{\mu\nu_1-\mu\nu_2-\nu_1^2+\nu_2^2}
(Q_{\nu_1}e^{\nu_1}-Q_{\nu_2}e^{\nu_2}-\frac{\nu_1^2-\nu_2^2}{\mu^2}Q_\mu e^\mu) \\
\end{aligned}
$$
is a \textit{decreasing} function of $\nu_2$, and the upper bound of
$e_1$ given in Eq.~\eqref{Decoy:e1Bound},
$$
\begin{aligned}
e_1^{U,\nu_1,\nu_2} &= \frac{E_{\nu_1} Q_{\nu_1}e^{\nu_1}-E_{\nu_2}
Q_{\nu_2}
e^{\nu_2}}{(\nu_1-\nu_2)Y_1^{L,\nu_1,\nu_2}} \\
\end{aligned}
$$
is an \textit{increasing} function of $\nu_2$, under conditions
Eq.~\eqref{Optnu2:Condition}. Here $Q_\mu$, $Q_{\nu_1}$,
$Q_{\nu_2}$, $E_\mu$, $E_{\nu_1}$ and $E_{\nu_2}$ are given by
Eqs.~\eqref{Model:Gain} and \eqref{Model:QBER}.

\noindent\textbf{Proof of the theorem} First we will prove
$Y_1^{L,\nu_1,\nu_2}$ is a decreasing function of $\nu_2$ and then
prove $e_1^{U,\nu_1,\nu_2}$ is an increasing function of $\nu_2$.

Define functions $G(\mu)$ and $J(\mu)$ as
$$
%\begin{equation}\label{Optnu2:Condistion}
\begin{aligned}
G(\mu) &= Q_\mu e^{\mu} \\
       &= (Y_0+1-e^{-\eta\mu})e^{\mu}\\
J(\mu) &= E_\mu Q_\mu e^{\mu} \\
       &= [e_0Y_0+e_{detector}(1-e^{-\eta\mu})]e^{\mu}.
\end{aligned}
%\end{equation}
$$
Take the first derivative of $G(\mu)$ and $J(\mu)$,
$$
%\begin{equation}\label{Optnu2:G1stDer}
\begin{aligned}
G'(\mu) &= Q_\mu e^{\mu} + \eta e^{(1-\eta)\mu}\\
%       &= (Y_0+1-e^{-\eta\mu})e^{\mu}.
J'(\mu) &= E_\mu Q_\mu e^{\mu} + \eta e_{detector} e^{(1-\eta)\mu},\\
\end{aligned}
%\end{equation}
$$
which are both increasing functions and $G'(\mu)\ge0$,
$J'(\mu)\ge0$. By mathematical induction, it is not difficult to
prove the following claim.

\noindent\textbf{Claim 1:} For any order derivative of $G(\mu)$ and
$J(\mu)$: $G^{(n)}(\mu)\ge0$ and $J^{(n)}(\mu)\ge0$ are increasing
functions.

\noindent\textbf{Some Useful Inequalities:}
With Claim 1 and the Taylor Series
of $G(\mu)$, we have
\begin{equation}\label{Optnu2:TaylorEx}
\begin{aligned}
G(\mu)&=\sum_{i=0}^{i=\infty}G^{(i)}(\mu)\frac{\mu^i}{i!}\\
&\ge\mu G'(\mu)
\end{aligned}
\end{equation}

According to Mean Value Theorem,
\begin{equation}\label{Optnu2:GMVT}
\begin{aligned}
\frac{G(\nu_1)-G(\nu_2)}{\nu_1-\nu_2} = G'(\nu_3) \\
\frac{J(\nu_1)-J(\nu_2)}{\nu_1-\nu_2} = J'(\nu_4)
\end{aligned}
\end{equation}
where $\nu_3,\nu_4\in[\nu_2,\nu_1]$. Because $G'(\mu)$ and $J'(\mu)$
are increasing functions, we can bound Eq.~\eqref{Optnu2:GMVT},
\begin{eqnarray}
%\begin{aligned}
G'(\nu_2) \le \frac{G(\nu_1)-G(\nu_2)}{\nu_1-\nu_2} \le G'(\nu_1) \label{Optnu2:G1st}\\
J'(\nu_2) \le \frac{J(\nu_1)-J(\nu_2)}{\nu_1-\nu_2} \le J'(\nu_1).
\label{Optnu2:J1st}
%\end{aligned}
\end{eqnarray}
Similarly,
\begin{equation}\label{Optnu2:G2nd}
\begin{aligned}
G''(\nu_2) \le \frac{G'(\nu_1)-G'(\nu_2)}{\nu_1-\nu_2} \le
G''(\nu_1).
\end{aligned}
\end{equation}

Define a function
$$
\begin{aligned}
F(\nu_2)&=\frac{1}{\mu-\nu_1-\nu_2}[Q_\mu
e^\mu-\frac{\mu}{\nu_1-\nu_2}(Q_{\nu_1}e^{\nu_1}-Q_{\nu_2}e^{\nu_2})]\\
&=\frac{1}{\mu-\nu_1-\nu_2}[G(\mu)-\frac{\mu}{\nu_1-\nu_2}(G({\nu_1})-G({\nu_2}))]
\end{aligned}
$$

\noindent\textbf{Claim 2:} The function $F(\nu_2)$ is an increasing
function of $\nu_2$, under the conditions given in
Eq.~\eqref{Optnu2:Condition}.

\noindent\textbf{Proof of Claim~2:} To determine if the function is
increasing or decreasing we will need the derivative.
\begin{equation}\label{Optnu2:Claim1Proof}
\begin{aligned}
F'(\nu_2)=&\frac{1}{(\mu-\nu_1-\nu_2)^2}[G(\mu)-\frac{\mu}{\nu_1-\nu_2}(G({\nu_1})-G({\nu_2}))] \\
&-
\frac{1}{\mu-\nu_1-\nu_2}\frac{\mu}{(\nu_1-\nu_2)^2}[G({\nu_1})-G({\nu_2})]
\\
&+ \frac{1}{\mu-\nu_1-\nu_2}\frac{\mu}{\nu_1-\nu_2}G'({\nu_2})
\\
\ge&\frac{1}{(\mu-\nu_1-\nu_2)^2}[G(\mu)-\mu G'({\nu_1})] \\
&- \frac{1}{\mu-\nu_1-\nu_2}\frac{\mu}{\nu_1-\nu_2}G'({\nu_1})
%\\
+ \frac{1}{\mu-\nu_1-\nu_2}\frac{\mu}{\nu_1-\nu_2}G'({\nu_2})
\\
\ge&\frac{1}{(\mu-\nu_1-\nu_2)^2}[\mu G'(\mu)-\mu G'({\nu_1+\nu_2})]
- \frac{\mu}{\mu-\nu_1-\nu_2}G''({\nu_1})
\\
\ge&\frac{\mu}{\mu-\nu_1-\nu_2}[G''({\nu_1+\nu_2}) - G''({\nu_1})]
\\
\ge& 0 \\
\end{aligned}
\end{equation}
Here, to prove the first inequality, we have made use of
Eq.~\eqref{Optnu2:G1st}; to prove the second inequality, we have
made use of Eq.~\eqref{Optnu2:TaylorEx}, \eqref{Optnu2:G2nd} and
Claim 1; to prove the third inequality, we have made use of
Eq.~\eqref{Optnu2:G2nd}; to prove the last inequality, we have made
use of Claim 1.

%Now, it comes to the main proof that Vacuum+Weak is optimal.
\textbf{Proof that $Y_1^{L,\nu_1,\nu_2}$ is a decreasing function.}
Re-write the lower bound of $Y_1$, in Eq.~\eqref{Decoy:Y1Bound},
\begin{equation}\label{Optnu2:Y1Bound}
\begin{aligned}
Y_1^{L,\nu_1,\nu_2} &= \frac{\mu}{\mu\nu_1-\mu\nu_2-\nu_1^2+\nu_2^2}
(Q_{\nu_1}e^{\nu_1}-Q_{\nu_2}e^{\nu_2}-\frac{\nu_1^2-\nu_2^2}{\mu^2}Q_\mu e^\mu) \\
&= \frac{\mu}{\mu\nu_1-\mu\nu_2-\nu_1^2+\nu_2^2}
(Q_{\nu_1}e^{\nu_1}-Q_{\nu_2}e^{\nu_2}) -
\frac{\mu}{\mu\nu_1-\mu\nu_2-\nu_1^2+\nu_2^2}\frac{\nu_1^2-\nu_2^2}{\mu^2}Q_\mu e^\mu \\
&= \frac{\mu}{\mu-\nu_1-\nu_2}
\frac{Q_{\nu_1}e^{\nu_1}-Q_{\nu_2}e^{\nu_2}}{\nu_1-\nu_2} -
\frac{\nu_1+\nu_2}{\mu-\nu_1-\nu_2}\frac{Q_\mu e^\mu}{\mu} \\
&= \frac{\mu}{\mu-\nu_1-\nu_2}
\frac{Q_{\nu_1}e^{\nu_1}-Q_{\nu_2}e^{\nu_2}}{\nu_1-\nu_2} -
(\frac{1}{\mu-\nu_1-\nu_2}-\frac{1}{\mu})Q_\mu e^\mu \\
&= \frac{1}{\mu}Q_\mu e^\mu - \frac{1}{\mu-\nu_1-\nu_2}[Q_\mu
e^\mu-\frac{\mu}{\nu_1-\nu_2}(Q_{\nu_1}e^{\nu_1}-Q_{\nu_2}e^{\nu_2})] \\
&= \frac{1}{\mu}Q_\mu e^\mu - F(\nu_2). \\
%&\le \frac{1}{\mu}Q_\mu e^\mu - \frac{1}{\mu-\nu_1}[Q_\mu
%e^\mu-\frac{\mu}{\nu_1-\nu_2}(Q_{\nu_1}e^{\nu_1}-Q_{\nu_2}e^{\nu_2})] \\
%&\le \frac{1}{\mu}Q_\mu e^\mu - \frac{1}{\mu-\nu_1}[Q_\mu
%e^\mu-\frac{\mu}{\nu_1}(Q_{\nu_1}e^{\nu_1}-Y_0)] \\
%&= \frac{\mu}{\mu\nu_1-\nu_1^2}(Q_{\nu_1} e^{\nu_1}-Q_\mu
%e^\mu\frac{\nu_1^2}{\mu^2}-Y_0) \\
%&\le \frac{\mu}{\mu\nu_1-\nu_1^2}(Q_{\nu_1} e^{\nu_1}-Q_\mu
%e^\mu\frac{\nu_1^2}{\mu^2}-\frac{\mu^2-\nu_1^2}{\mu^2}Y_0) \\
%&= Y_1^{L,\nu_1,0}. \\
\end{aligned}
\end{equation}
With Claim 2, we show that $Y_1^{L,\nu_1,\nu_2}$ is a decreasing
function of $\nu_2$.

Define a function
$$
\begin{aligned}
K(\nu_2)&=\frac{E_{\nu_1} Q_{\nu_1}e^{\nu_1}-E_{\nu_2} Q_{\nu_2}
e^{\nu_2}}{\nu_1-\nu_2} \\
&=\frac{J({\nu_1})-J({\nu_2})}{\nu_1-\nu_2}
\end{aligned}
$$

\noindent\textbf{Claim 3:} function $K(\nu_2)$ is an increasing
function with $\nu_2$.

\noindent\textbf{Proof:} to determine if the function is increasing
or decreasing we will need the derivative.
\begin{equation}\label{Optnu2:Claim2Proof}
\begin{aligned}
K'(\nu_2) &=
\frac{J({\nu_1})-J({\nu_2})}{(\nu_1-\nu_2)^2}-\frac{J'({\nu_2})}{\nu_1-\nu_2}
\\
&\ge \frac{J'({\nu_2})}{\nu_1-\nu_2}-\frac{J'({\nu_2})}{\nu_1-\nu_2}
\\ &= 0,
\end{aligned}
\end{equation}
where the first inequality is due to Eq.~\eqref{Optnu2:J1st}.

\textbf{Proof that $e_1^{U,\nu_1,\nu_2}$ is an increasing function.}
Reform the lower bound of $e_1$, in Eq.~\eqref{Decoy:e1Bound},
\begin{equation}\label{Optnu2:e1Bound}
\begin{aligned}
e_1^{U,\nu_1,\nu_2} &= \frac{E_{\nu_1} Q_{\nu_1}e^{\nu_1}-E_{\nu_2}
Q_{\nu_2}
e^{\nu_2}}{(\nu_1-\nu_2)Y_1^{L,\nu_1,\nu_2}} \\
&= \frac{K(\nu_2)}{Y_1^{L,\nu_1,\nu_2}}
%&\ge \frac{E_{\nu_1} Q_{\nu_1}e^{\nu_1}-e_0Y_0}{\nu_1Y_1^{L,\nu_1,\nu_2}} \\
%&\ge \frac{E_{\nu_1} Q_{\nu_1}e^{\nu_1}-e_0Y_0}{\nu_1Y_1^{L,\nu_1,0}} \\
%&= e_1^{U,\nu_1,0}
\end{aligned}
\end{equation}
With Claim 3 and decreasing function of $Y_1^{L,\nu_1,\nu_2}$, we
show that $e_1^{U,\nu_1,\nu_2}$ is an increasing function of
$\nu_2$.

%In summary, with Eqs.~\eqref{Optnu2:Y1Bound} and
%\eqref{Optnu2:e1Bound}, we conclude that Vacuum+Weak ($\nu_2=0$)
%gives optimal estimations of $Y_1$ and $e_1$. More generally, for
%two-decoy-state, the smaller $\nu_2$ is, the better the estimations
%of $Y_1$ and $e_1$ are.

In summary, we have proved the theorem.

\end{appendix}


\begin{thebibliography}{99}
\bibitem{gain} Here, the gain of a particular type of signals
is defined to be the fraction of detection events by Bob that is due to that
particular type of signals.
\end{thebibliography}
\end{document}